\documentclass[a4paper,11pt]{article}

\usepackage{cite}

\usepackage{fullpage,setspace,hyperref,comment,caption}
\usepackage[font=footnotesize,labelfont=bf,margin=1cm]{caption}
\usepackage{cite} 
\usepackage{color}
\usepackage{amsfonts}
\usepackage{amsmath}
\usepackage{amssymb}
\usepackage{todonotes} 
\usepackage[utf8]{inputenc}
\usepackage{multirow}					
\usepackage{multicol}
\usepackage{pdflscape}
\usepackage{afterpage}

\numberwithin{equation}{section}

\newcommand{\beq}{\begin{equation}}
\newcommand{\eeq}{\end{equation}}
\newcommand{\be}{\begin{equation}}
\newcommand{\ee}{\end{equation}}

\newcommand{\dd}{\mathrm{d}}		
			
\newcommand{\cH}{\mathcal{H}}			
\newcommand{\MM}{\mathcal{M}}
\newcommand{\gM}{\mathcal{M}}
\newcommand{\LL}{\mathcal{L}}			

\newcommand{\hR}{\hat{R}}
\newcommand{\bz}{\bar{z}}

\newcommand{\hmu}{{\hat{\mu}}}
\newcommand{\hnu}{{\hat{\nu}}}
\newcommand{\ha}{\hat{a}}
\newcommand{\hrho}{\hat{\rho}}
\newcommand{\hlambda}{\hat{\lambda}}
\newcommand{\bm}{\bar{m}}

\newcommand{\gL}{\LL}

\newcommand{\1}{{\mu_1}}
\newcommand{\2}{{\mu_2}}
\newcommand{\3}{{\mu_3}}
\newcommand{\4}{{\mu_4}}
\newcommand{\5}{{\mu_5}}
\newcommand{\6}{{\mu_6}}
\newcommand{\7}{{\mu_7}}
\newcommand{\8}{{\mu_8}}

\newcommand{\mt}{{\mu_{10}}}

\newcommand{\p}{\bullet}
\newcommand{\pl}{\!\!\p\!}

\newcommand{\hpartial}{\hat{\partial}}
\newcommand{\hd}{\hpartial}

\newcommand{\D}{D}

\newcommand{\TAw}{\TA(1/7)}
\newcommand{\TBw}{\TB(2/7)}
\newcommand{\TCw}{\TC(3/7)}
\newcommand{\TDw}{\TD(4/7)}
\newcommand{\TEw}{\TE(5/7)}
\newcommand{\TFw}{\TF(6/7)}
\newcommand{\TSw}{\TS(1)}
\newcommand{\TA}{\mathbb{A}}
\newcommand{\TB}{\mathbb{B}}
\newcommand{\TC}{\mathbb{C}}
\newcommand{\TD}{\mathbb{D}}
\newcommand{\TE}{\mathbb{E}}
\newcommand{\TF}{\mathbb{F}}
\newcommand{\TS}{\mathbb{S}}

\newcommand\Tstrut{\rule{0pt}{3ex}}         
\newcommand\Bstrut{\rule[-1.3ex]{0pt}{0pt}}   

\newcommand{\Aa}{A}
\newcommand{\Ab}{B}
\newcommand{\Ac}{C}
\newcommand{\Ad}{D}
\newcommand{\Ae}{E}
\newcommand{\Af}{F}

\newcommand{\Fa}{\mathcal{F}}
\newcommand{\Fb}{\mathcal{H}}
\newcommand{\Fc}{\mathcal{J}}
\newcommand{\Fd}{\mathcal{K}}
\newcommand{\Fe}{\mathcal{L}}
\newcommand{\bF}{\bar{F}}
\newcommand{\bB}{\bar{C}}
\newcommand{\bC}{\bar{C}}

\newcommand{\bA}{\bar{C}}

\newcommand{\G}{\mathrm{SL}(2) \times \mathbb{R}^+}
\newcommand{\Edd}{E_{d(d)}}

\newcommand{\anom}{\hat{\Delta}}

\begin{document}

\begin{titlepage}
\vfill

\begin{flushright}
QMUL-PH-15-26 \\
LMU-ASC 79/15
\end{flushright}

\vfill

\begin{center}
   \baselineskip=16pt
   	{\Large \bf An Action for F-theory: \\ $\mathrm{SL}(2) \times \mathbb{R}^+$ Exceptional Field Theory}
   	\vskip 2cm
   	{\sc		David S. Berman$^*$\footnote{\tt d.s.berman@qmul.ac.uk}, Chris D. A. Blair$^\dagger$\footnote{\tt cblair@vub.ac.be}, \\ Emanuel Malek$^\sharp$\footnote{\tt e.malek@lmu.de}, Felix J. Rudolph$^*$\footnote{\tt f.j.rudolph@qmul.ac.uk}}
	\vskip .6cm
    {\small \it $*$ Queen Mary University of London, Centre for Research in String Theory, \\
             School of Physics, Mile End Road, London, E1 4NS, England \\ \ \\
            $\dagger$ Theoretische Natuurkunde, Vrije Universiteit Brussel, and the International Solvay Institutes, \\ 
            Pleinlaan 2, B-1050 Brussels, Belgium \\ \ \\
            $\sharp$ Arnold Sommerfeld Center for Theoretical Physics, Department f\"ur Physik, \\ Ludwig-Maximilians-Universit\"at M\"unchen, Theresienstra{\ss}e 37, 80333 M\"unchen, Germany} \\ 
	\vskip 2cm
\end{center}

\begin{abstract}

We construct the 12-dimensional exceptional field theory associated to the group $\G$. Demanding the closure of the algebra of local symmetries leads to a constraint, known as the section condition, that must be imposed on all fields. This constraint has two inequivalent solutions, one giving rise to 11-dimensional supergravity and the other leading to F-theory. Thus $\G$ exceptional field theory contains both F-theory and M-theory in a single 12-dimensional formalism.

\end{abstract}

\vfill

\setcounter{footnote}{0}
\end{titlepage}

\tableofcontents

\section{Introduction}

It is 20 years since the study of non-perturbative string dynamics \cite{Witten:1995ex} and U-duality \cite{Hull:1994ys} led to the idea of M-theory. From its inception the low energy effective description of M-theory was known to be 11-dimensional supergravity with the coupling of the Type IIA string promoted to the radius of the eleventh dimension. 

The natural extension of this idea to the type IIB string gave rise to F-theory \cite{Vafa:1996xn} where the complex coupling in the IIB theory is taken to have its origin in the complex modulus of a torus fibred over the usual ten dimensions of the type IIB string theory. Thus by definition, F-theory is the 12-dimensional lift of the type IIB string theory. The status of this 12-dimensional theory has been somewhat different to that of its IIA spouse with no direct 12-dimensional description in terms of an action and fields that reduce to the IIB theory. Indeed, there is no 12-dimensional supergravity and thus no limit in which the 12-dimensions can be taken to be ``large'', unlike in the M-theory case. The emphasis has thus largely been on using algebraic geometry to describe F-theory compactifications \cite{Morrison:1996na, Morrison:1996pp} such that now F-theory is synonymous with the study of elliptically fibred Calabi-Yau manifolds. 

The complex coupling of IIB theory is naturally acted on by an $\mathrm{SL}(2)$ S-duality, which is a symmetry of the theory. F-theory provides a geometric interpretation of this duality. An idea which was raised swiftly after the introduction of F-theory was whether there exist similar geometrisations of the U-duality symmetries which one finds after descending in dimension \cite{Kumar:1996zx}. Here, one would hope to be able to associate the scalars of compactified supergravity with the moduli of some auxiliary geometric space. However, the scope for doing so turns out to be somewhat limited.

More recently, the idea of reimagining duality in a geometric origin has resurfaced in the form of double field theory (DFT) and exceptional field theory (EFT). Perhaps the key development that allows for the construction of these theories is that the new geometry is not a conventional one, but an ``extended geometry'' \cite{Coimbra:2011ky,Coimbra:2012af} based on the idea of ``generalised geometry'' \cite{Hitchin:2004ut, Gualtieri:2003dx}. For instance, a key role is played by a ``generalised metric'', in place of the torus modulus, and one introduces an extended space with a novel ``generalised diffeomorphism'' symmetry. 

In this paper, we aim to use these innovations to construct a 12-dimensional theory which provides a local action for F-theory. 

Double field theory \cite{
Hull:2009mi, Hull:2009zb, Hohm:2010jy, Hohm:2010pp
} - which itself has roots older still than F-theory \cite{Duff:1989tf, Tseytlin:1990nb, Tseytlin:1990va, Siegel:1993xq,Siegel:1993th} - provides an $O(d,d)$ symmetry which allows one to think of the $O(d,d)$ T-duality symmetry in a generalised geometric manner. The bosonic degrees of freedom of $\mathcal{N}=1$ supergravity are described by a generalised metric which transforms under and is itself an element of $O(d,d)$. The key difference to generalised geometry is the addition of novel coordinates so that the ``doubled coordinates'' transform under the fundamental representation of $O(d,d)$.\footnote{The introduction of extra coordinates is also important for the description of so-called non-geometric backgrounds\cite{Hohm:2013bwa}.} 
This is a natural construction from closed string field theory \cite{Kugo:1992md}. 

The extension of the ideas of double field theory to the exceptional sequence of groups $\Edd$, which are associated with U-duality, is called exceptional field theory \cite{Hillmann:2009ci, Berman:2010is, Berman:2011cg, Berman:2011jh, Hohm:2013pua}. Originally developed in terms of an extended geometry \cite{Hillmann:2009ci, Berman:2010is, Berman:2011cg, Berman:2011jh, Park:2013gaj, Aldazabal:2013mya, Blair:2013gqa, Blair:2014zba} geometrising the $\Edd$ groups acting on a truncated $d$-dimensional theory, there is now a systematic method for constructing EFT in a form which is fully equivalent to the whole 11- or 10-dimensional supergravities \cite{Hohm:2013pua}. So far EFTs have been developed for $3\leq d\leq8$: $E_8$ \cite{Hohm:2014fxa}, $E_7$ \cite{Hohm:2013uia}, $E_6$ \cite{Hohm:2013vpa}, $\mathrm{SO}(5,5)$ \cite{Abzalov:2015ega}, $\mathrm{SL}(5)$ \cite{Musaev:2015ces} and $\mathrm{SL}(3) \times \mathrm{SL}(2)$ \cite{Hohm:2015xna} theory, including the supersymmetrisation of the $E_7$ and $E_6$ cases \cite{Godazgar:2014nqa,Musaev:2014lna}. The closely-related generalised geometry construction \cite{Coimbra:2011ky,Coimbra:2012af} is supersymmetric for the exceptional groups considered, $E_4$ up to $E_8$.
In addition, fairly complete constructions have been developed for the general field content of these theories \cite{Cederwall:2013naa,Wang:2015hca}. Some recent reviews - predating the construction of the full EFTs - are \cite{Aldazabal:2013sca, Berman:2013eva, Hohm:2013bwa}. 

We shall now give an executive summary of the EFT construction for the purposes of this general introduction, following the original line of thinking on the subject:
\begin{itemize}
\item Let us consider ourselves as starting with 11-dimensional supergravity.
\item We decompose the 11-dimensional diffeomorphism group $\mathrm{GL}(11) \longrightarrow \mathrm{GL}(11-d) \times \mathrm{GL}(d)$. 
\item We promote the $\mathrm{SL}(d) \subset \mathrm{GL}(d)$ symmetry to $\Edd$ by rewriting the bosonic degrees of freedom in terms of objects which fall naturally into representations of $\Edd$ (in order to do so, we may need to perform various dualisations).
\item We now take the step of introducing $N$ new coordinates such that the $d$ original coordinates and the $N$ extra (or ``dual'') coordinates fit into a certain representation of $\Edd$.
\item Our fields - which are already in representations of $\Edd$ - are now taken to depend on all the coordinates. 
\item The local symmetries of the theory are likewise rewritten in such a manner that diffeomorphisms and gauge transformations unify into ``generalised diffeomorphisms''.
\item Invariance under these local symmetries may be used to fix the EFT action in the total $(11+N)$-dimensional space.
\item For consistency the theory must be supplemented with a constraint, quadratic in derivatives, known as the {\it{section condition}}. Inequivalent solutions (in the sense that they cannot be mapped into each other by an $\Edd$ transformation) of this constraint will impose a dimensional reduction of the $(11+N)$-dimensional theory down to eleven or ten dimensions. The reduction down to eleven we call the M-theory section and the reduction down to ten the IIB section for reasons that will become obvious. Note that although one may motivate and begin the construction by starting in either eleven dimensions, or ten, the full theory will automatically contain both M-theory and IIB subsectors \cite{Hohm:2013vpa, Blair:2013gqa}
\end{itemize}

In this paper we give the result of following this procedure to construct the EFT relevant to F-theory. The result will be a 12-dimensional theory which may be reduced either to the maximal supergravity in eleven dimensions or type IIB supergravity in ten. This theory is manifestly invariant under a $\G$ symmetry. The $\mathrm{SL}(2)$ part of this is easily identified with the S-duality of type IIB and after imposing the section condition one can see the emergence of a fibration structure as considered in F-theory. A different attempt at realising a twelve-dimensional gravity description can be found in \cite{Choi:2015gia}. There have been other attempts at making U-dualities manifest under the name of ``F-theory'' \cite{Linch:2015lwa,Linch:2015fya,Linch:2015qva,Linch:2015fca,Siegel:2016dek}. These should not be confused with our approach which is based on the more traditional interpretation of F-theory as a 12-dimensional framework making the $\mathrm{SL}(2)$ S-duality of type IIB manifest.

We note that the construction is a little different to the cases considered before in that the extended space that is introduced is too small to include any effects from geometrising gauge field potentials such as the three-form of 11-dimensional supergravity. In some sense, one may think of it as the minimal EFT. As such, it provides the easiest way of seeing exactly how F-theory fits in to the general EFT framework. By constructing the theory our goal is to bring into sharp focus the points of comparison between the EFT construction and what is normally thought of as F-theory. It will be of interest to embed this viewpoint into the higher rank groups already studied. 

In this paper, after describing the action and symmetries of the $\G$ EFT in section \ref{sec:theory}, we begin the process of analysing how it relates to F-theory. We can precisely show the identification of the fields of the $\G$ EFT with those of 11-dimensional supergravity and 10-dimensional IIB supergravity, which we do in section \ref{sec:supergravity}. After this, in section \ref{sec:Frel} we discuss in general terms how one may view the IIB section - where there is no dependence on two coordinates - as an F-theory description. The dynamics of the extended space are encoded in a generalised metric whose degrees of freedom are precisely such as to allow us to recover the usual sort of torus fibration familiar in F-theory. 

We also discuss in detail how one sees M-theory/F-theory duality in the EFT framework. One can work out the precise maps between the M-theory and F-theory pictures by comparing the identifications of the EFT fields in the different sections. In this way one recovers the usual relationships relating 11-dimensional supergravity on a torus to IIB string theory on a circle in nine dimensions \cite{Schwarz:1995jq}. We carry this out for a selection of familiar brane solutions in section \ref{sec:solutions}, paying attention to the different features that are seen in EFT depending on whether the brane wraps part of the extended space or not. For instance, 1/2-BPS brane solutions may be seen as simple wave/monopole type solutions in the EFT \cite{Berman:2014hna}. 

At the same time, we stress that the EFT construction is an extension of the usual duality relationships. From the point of view of the M-theory section, we introduce one additional dual coordinate $y^s$, which can be thought of as being conjugate to a membrane wrapping mode when we have a background of the form $M_9 \times T^2$. Alternatively, the IIB section sees two dual coordinates $y^\alpha$ related to winding modes of the F1 and D1 in the direction $y^s$. In EFT we treat all these modes on an equal footing - they are related by the symmetry $\G$.\footnote{Actually, in this case $y^s$ is a singlet of the $\mathrm{SL}(2)$ part of the group while $y^\alpha$ is a doublet.}

We hope that by presenting the 12-dimensional $\G$ exceptional field theory we can begin to properly connect the EFT constructions to F-theory. We conclude with a discussion of possible directions in which one can use the EFT formalism to address aspects of F-theory, and comment on the outlook to higher dimensional duality groups.

\section{The $\mathrm{SL}(2) \times \mathbb{R}^+$ exceptional field theory} 
\label{sec:theory}

In this section, we describe the general features of the theory. First, we describe the setup in terms of an extended space and the field content. Then we shall give the form of the action. After this, we will briefly discuss the symmetries of the theory, which can be used to determine the action. Our goal is to provide an introduction to the main features. Thus, we reserve the technical details of the construction to the appendices. 

In this paper will restrict ourselves to only the bosonic sector of the theory. We expect that supersymmetrisation will follow the form of other supersymmetrised EFTs \cite{Godazgar:2014nqa,Musaev:2014lna}.

\subsection{The theory and its action}

The theory we will describe in this paper may be thought of as a 12-dimensional theory with a $9+3$ split of the coordinates, so that we have
\begin{itemize}
\item nine ``external'' coordinates, $x^\mu$,
\item three ``extended'' coordinates, $Y^M$ that live in the $\mathbf{2}_{1} \oplus \mathbf{1}_{-1}$ reducible representation of $\G$ (where the subscripts denote the weights under the $\mathbb{R}^+$ factor). To reflect the reducibility of the representation we further decompose the coordinates $Y^M = ( y^\alpha, y^s)$ where $\alpha=1,2$ transforms in the fundamental of $\mathrm{SL}(2)$, and $s$ stands, appropriately as we will see, for ``singlet'' or ``string''. \footnote{The reducibility of the coordinate representation is not a feature of higher rank duality groups.}
\end{itemize}
The fields and symmetry transformation parameters of the theory can in principle depend on all of these coordinates. However, as always happens in exceptional field theory and double field theory, there is a consistency condition which reduces the dependence on the extended coordinates. This condition is usually implemented as the section condition, which directly imposes that that the fields cannot depend on all extended coordinates. In our case, it takes the form
\be
\partial_\alpha \partial_s = 0 \,,
\label{eq:constraint}
\ee
with the derivatives to be thought of as acting on any field or pair of fields, so that we require both $\partial_\alpha \partial_s \mathcal{O} =0$ and $\partial_\alpha \mathcal{O}_1 \partial_s \mathcal{O}_2 + \partial_\alpha \mathcal{O}_2 \partial_s \mathcal{O}_1 = 0$. The origin of the section condition is the requirement that the algebra of symmetries closes, which we will review in the next subsection.\footnote{In fact, one can alternatively use a generalised Scherk-Schwarz ansatz to obtain a set of weaker conditions that replace the above section condition with a set of constraints on twist matrices which in principle depend on all the extended coordinates \cite{Aldazabal:2011nj,Geissbuhler:2011mx,Grana:2012rr,Berman:2012uy}. This would be an interesting possibility to apply here, but we do not pursue this in the present paper.}

The field content of the theory is as follows. The metric-like degrees of freedom are
\begin{itemize}
\item an ``external'' metric, $g_{\mu \nu}$, 
\item a generalised metric, $\gM_{MN}$ which parametrises the coset $(\G)/\mathrm{SO}(2)$. (From the perspective of the ``external'' nine dimensions, this metric will correspond to the scalar degrees of freedom.) The reducibility of the $Y^M$ coordinates implies that the generalised metric is reducible and thus may be decomposed as, \be \gM_{MN}=\gM_{\alpha \beta} \oplus \gM_{ss} \, . \ee
\end{itemize}
The coset $(\G)/\mathrm{SO}(2)$ implies we have just three degrees of freedom described by the generalised metric. This means that $\gM_{ss}$ must be related to $\det \gM_{\alpha \beta}$. One can thus define $\gM_{\alpha \beta}$ such that
\be
\cH_{\alpha \beta} \equiv (\gM_{ss})^{3/4} \gM_{\alpha \beta} \,, \label{eq:unitdetgenmetric}
\ee
has unit determinant. The rescaled metric $\cH_{\alpha \beta}$ and $\gM_{ss}$ can then be used as the independent degrees of freedom when constructing the theory. This unit determinant matrix $\cH_{\alpha \beta}$ will then appear naturally in the IIB/F-theory description. 

In addition, we have a hierarchy of gauge fields, similar to the tensor hierarchy of gauged supergravities \cite{deWit:2005hv,deWit:2008ta}. These are form fields with respect to the external directions, and also transform in different representations of the duality group:
\be
\begin{array}{|c|ll|ll|} \hline
\mathrm{Representation} & & \mathrm{Gauge\ potential} & & \mathrm{Field}\,\, \mathrm{strength} \\ \hline
\mathbf{2}_1 \oplus \mathbf{1}_{-1} & & \Aa_\mu{}^M & & \Fa_{\mu \nu}{}^M \\
\mathbf{2}_0 & & \Ab_{\mu \nu}{}^{\alpha s} & & \Fb_{\mu \nu \rho}{}^{\alpha s} \\ 
\mathbf{1}_1 & & \Ac_{\mu \nu \rho}{}^{[\alpha \beta] s} & & \Fc_{\mu \nu \rho \sigma}{}^{[\alpha \beta] s} \\
\mathbf{1}_0 & & \Ad_{\mu \nu \rho \sigma}{}^{[\alpha \beta] ss } & & \Fd_{\mu \nu \rho \sigma \lambda}{}^{[\alpha \beta]ss} \\
\mathbf{2}_1 & & \Ae_{\mu \nu \rho \sigma \kappa}{}^{\gamma [\alpha \beta] ss } & & \Fe_{\mu \nu \rho \sigma \kappa\lambda}{}^{\gamma [\alpha \beta]ss} \\
\mathbf{2}_0 \oplus \mathbf{1}_2 & & \Af_{\mu \nu \rho \sigma \kappa \lambda}{}^{M} & & \mathrm{not}\,\,\mathrm{needed} \\  \hline
\end{array}
\label{eq:forms}
\ee
These fields also transform under ``generalised diffeomorphisms'' and ``external diffeomorphisms'' which we describe in the next subsection, as well as various gauge symmetries of the tensor hierarchy which we describe in Appendix \ref{sec:tensorhier}. The field strengths are defined such that the fields transform covariantly under generalised diffeomorphisms, i.e. according to their index structure and the rules given in the following subsection, and are gauge invariant under a hierarchy of interrelated gauge transformations given in detail in \ref{sec:tensorhier}. The expressions for the field strengths are schematically
\be
\begin{split} 
\mathcal{F}_{\mu \nu}{} & = 2 \partial_{[\mu} A_{\nu]} +  \dots + \hd\Ab_{\1\2}\, \\
\Fb_{\1\2\3} &= 3\D_{[\1}\Ab_{\2\3]} +  \dots + \hd\Ac_{\1\2\3} \,, \\
\Fc_{\1\ldots\1\4} &= 4\D_{[\1}\Ac_{\2\3\4]}  + \dots + \hd\Ad_{\1\ldots\4} \,, \\
\Fd_{\1\ldots\5} &= 5\D_{[\1}\Ad_{\2\ldots\5]} + \dots + \hd\Ae_{\1\ldots\5} \,, \\
\Fe_{\1\ldots\6} &= 6\D_{[\1}\Ae_{\2\ldots\6]}  + \dots + \hd\Af_{\1\ldots\6} \,,
\end{split} 
\ee
where for the $p$-form field strengths the terms indicated by dots involve quadratic and higher order of field strengths.
We also see that there is always a linear term, shown, of the gauge field at next order, under a particular nilpotent derivative $\hd$ defined in \ref{sec:cartancalc}. The derivative, $D_\mu$ that appears is a covariant derivative for the generalised diffeomorphisms, as described in the next section.
The detailed definitions of the field strengths are also in \ref{sec:tensorhier}. 

Crucially, the presence of the two kinds of diffeomorphism symmetries may be used to fix the action up to total derivatives. The details of this calculation are in Appendix \ref{sec:detact}. The resulting general form of the action, which is common to all exceptional field theories is given schematically as follows, 
\be
S = \int \dd^{9} x \dd^3 Y \sqrt{g} \left( \hR + 
\LL_{skin} 
+ \LL_{gkin} 
+ \frac{1}{\sqrt{g}} \LL_{top} 
+ V 
\right) \,.
\label{eq:S} 
\ee
The constituent parts are (omitting total derivatives\footnote{In the quantum theory the total derivatives are important and EFT will have a natural boundary term given by an $\Edd$ covariantisation of the Gibbons-Hawking term \cite{Berman:2011kg}.}):
\begin{itemize}
\item the ``covariantised'' external Ricci scalar, $\hR$, which is 
\be
\hR = 
\frac{1}{4} g^{\mu \nu} D_\mu g_{\rho \sigma} D_\nu g^{\rho \sigma} 
- \frac{1}{2} g^{\mu \nu} D_\mu g^{\rho \sigma} D_\rho g_{\nu \sigma} 
+ \frac{1}{4} g^{\mu \nu} D_\mu \ln g D_\nu \ln g + \frac{1}{2}  D_\mu \ln g D_\nu g^{\mu \nu} \,.
\ee 
\item a kinetic term for the generalised metric
\be
\mathcal{L}_{skin}
= -\frac{7}{32} g^{\mu \nu} D_\mu \ln \gM_{ss} D_\nu \ln \gM_{ss} 
+ \frac{1}{4}  g^{\mu \nu} D_\mu \cH_{\alpha \beta} D_\nu \cH^{\alpha \beta} \, ,
\ee

\item kinetic terms for the gauge fields
\be
\begin{split}
\mathcal{L}_{gkin} & = 
- \frac{1}{2\cdot 2!} \gM_{MN} \mathcal{F}_{\mu \nu}{}^M \mathcal{F}^{\mu \nu N} 
 - \frac{1}{2\cdot 3!} \gM_{\alpha \beta} \gM_{ss} \mathcal{H}_{\mu \nu \rho}{}^{\alpha s} \mathcal{H}^{\mu \nu \rho \beta s} 
\\ & \qquad - \frac{1}{2\cdot 2! 4!} \gM_{ss} \gM_{\alpha \gamma} \gM_{\beta \delta} \mathcal{J}_{\mu \nu \rho \sigma}{}^{[\alpha \beta] s} \mathcal{J}^{\mu \nu \rho \sigma [\gamma \delta]s} 
\,.
\end{split}
\ee
We do not include kinetic terms for all the form fields appearing in \eqref{eq:forms}. As a result, not all the forms are dynamical. We will discuss the consequences of this below. 
\item a topological or Chern-Simons like term which is not manifestly gauge invariant in 9+3 dimensions. In a standard manner however we may write this term in a manifestly gauge invariant manner in 10+3 dimensions as
\begin{equation}
 \begin{split}
  S_{top} &= \kappa \int d^{10}x\, d^3Y\, \varepsilon^{\1\ldots\mt}  \frac{1}{4}  \epsilon_{\alpha \beta} \epsilon_{\gamma \delta}  \left[ 
  \frac{1}{5}  \partial_s \Fd_{\mu_1 \dots \mu_5}{}^{\alpha \beta ss} \Fd_{\mu_6 \dots \mu_{10}}{}^{\gamma \delta ss} \right. \\ 
  & \qquad \left. - \frac{5}{2} \Fa_{\mu_1 \mu_2}{}^s \Fc_{\3\ldots\6}{}^{\alpha \beta s} \Fc_{\7\ldots\mt}{}^{\gamma \delta}  \right. \\
    & \qquad \left. + \frac{10}{3}2 \Fb_{\mu_1 \dots \mu_3}{}^{\alpha s}\Fb_{\4\ldots\6}{}^{\beta s} \Fc_{\7 \ldots \mt}{}^{\gamma \delta}  \right] \,.
 \end{split} \label{eq:ToptTerm}
\end{equation}
The index $\mu$ is treated to an abuse of notation where it is simultaneously 10- and 9-dimensional. (This extra dimension is purely a notational convenience and is unrelated to the extra coordinates present in $Y^M$.) The above term is such that its variation is a total derivative and so can be written again in the correct number of dimensions. For further discussion of this term, including an ``index-free'' description, see Appendix \ref{sec:topterm}. The overall coefficient $\kappa$ is found to be $\kappa = +\frac{1}{5!48}$. 
\item a scalar potential
\be
\begin{split}
V & = 
\frac{1}{4} \gM^{ss}\left( 
 \partial_s \cH^{\alpha \beta} \partial_s \cH_{\alpha \beta} 
 + \partial_s g^{\mu \nu} \partial_s g_{\mu \nu} + \partial_s \ln g \partial_s \ln g 
 \right) 
 \\ & 
 +  \frac{9}{32}  \gM^{ss} \partial_s \ln \gM_{ss} \partial_s \ln \gM_{ss}
- \frac{1}{2}  \gM^{ss} \partial_s \ln \gM_{ss} \partial_s \ln g 
\\ &  +
\gM_{ss}^{3/4} \Bigg[ 
\frac{1}{4} \cH^{\alpha \beta} \partial_\alpha \cH^{\gamma \delta} \partial_\beta \cH_{\gamma \delta} 
+ \frac{1}{2} \cH^{\alpha \beta} \partial_\alpha \cH^{\gamma \delta} \partial_\gamma \cH_{\delta \beta} 
+ \partial_\alpha \cH^{\alpha \beta} \partial_\beta \ln \left(  g^{1/2} \gM_{ss}^{3/4}  \right)  
\\ & 
+ \frac{1}{4} \cH^{\alpha \beta} \left( \partial_\alpha g^{\mu \nu} \partial_\beta g_{\mu \nu} 
+   \partial_\alpha \ln g \partial_\beta \ln g
+ \frac{1}{4} \partial_\alpha \ln \gM_{ss} \partial_\beta \ln \gM_{ss}  + \frac{1}{2} \partial_\alpha \ln g \partial_\beta \ln \gM_{ss} 
 \right) 
\Bigg] \,.
\end{split}
\ee
\end{itemize}

This theory expresses the dynamics of 11-dimensional supergravity and 10-dimensional type IIB supergravity in a duality covariant way. In order to do so, we have actually increased the numbers of degrees of freedom by simultaneously treating fields and their electromagnetic duals on the same footing. This can be seen in the collection of form fields in \eqref{eq:forms}. For instance, although 11-dimensional supergravity contains only a three-form, here we have additional higher rank forms which can be thought of as corresponding to the six-form field dual to the three-form. 

The action for the theory deals with this by not including kinetic terms for all the gauge fields. The field strength $\Fd_{\mu\nu\rho\sigma\kappa}$ of the gauge field $D_{\mu \nu \rho \sigma}$ only appears in the topological term \eqref{eq:ToptTerm}. The field $D_{\mu\nu\rho\sigma}$ in fact also appears in the definition of the gauge field $\Fc_{\mu \nu \rho}$, under a $\partial_M$ derivative. One can show that the equation of motion for this field is 
\be
\partial_s \left( \frac{\kappa}{2} \epsilon^{\mu_1 \dots \mu_9} \epsilon_{\alpha \beta} 
\epsilon_{\gamma \delta} \Fd_{\mu_5 \dots \mu_9}{}^{\gamma \delta ss} 
-  e \frac{1}{48} \gM_{ss}  \gM_{\alpha \gamma} \gM_{\beta \delta} \Fc^{\mu_1 \dots \mu_4 \gamma \delta s} \right) = 0  \,.
\label{eq:Deom} 
\ee
The expression in the brackets should be imposed as a duality relation relating the field strength $\Fd_{\mu \nu \rho \sigma \lambda}$ to $\Fc_{\mu \nu \rho \sigma}$, and hence removing seemingly extra degrees of freedom carried in the gauge fields which are actually just the dualisations of physical degrees of freedom. The above relation is quite important -- for instance the proof that the EFT action is invariant under diffeomorphisms is only obeyed if it is satisfied. 

As for the remaining two gauge fields, the equation of motion following from varying with respect to $E_{\mu \nu \rho \sigma \kappa}$ is trivially satisfied (it only appears in the field strength $\Fd_{\mu \nu \rho \sigma \kappa}$), while $F_{\mu \nu \rho \sigma \kappa \lambda}$ is entirely absent from the action. 

\subsection{Local and global symmetries} 

It is clear that the above action has a manifest invariance under a global $\G$ symmetry, acting on the indices $\alpha,s$ in an obvious way. In addition, the exceptional field theory is invariant under a set of local symmetries.

Alongside the introduction of the extended coordinates $Y^M$ one constructs so called ``generalised diffeomorphisms". In the higher rank groups, these give a unified description of ordinary diffeomorphisms together with the $p$-form gauge transformations. Although the group $\G$ is too small for the $p$-form gauge transformations to play a role here, the generalised diffeomorphisms provide a combined description of part of the ordinary local symmetries of IIB and 11-dimensional supergravity.

The generalised diffeomorphisms, generated by a generalised vector $\Lambda^M$, act as a local $\G$ action, called the generalised Lie derivative $\gL_\Lambda$. These act on a vector, $V^M$ of weight $\lambda_V$ in a form which looks like the usual Lie derivative plus a modification involving the so-called ``Y-tensor''
\be
\delta_\Lambda V^M \equiv
\mathcal{L}_\Lambda V^M 
= \Lambda^N \partial_N V^M 
- V^N \partial_N \Lambda^M 
+ Y^{MN}{}_{PQ} \partial_N \Lambda^P V^Q 
+ ( \lambda_V + \omega) \partial_N \Lambda^N V^M \,.
\label{eq:gld}
\ee
This modification is universal for the generalised Lie derivatives of all exceptional field theories \cite{Berman:2012vc} and is built from the invariant tensors of the duality group. This universal formulation of the Lie derivative generalises the construction of DFT \cite{Siegel:1993th,Hull:2009zb}, as well as that given for the $\mathrm{SL}(5)$ EFT \cite{Berman:2011cg}, and indeed was first given in the context of generalised geometry \cite{Coimbra:2011ky}, where the derivatives with respect to the extra coordinates are set to zero. In the case of $\G$ \cite{Wang:2015hca}, the $Y$-tensor is symmetric on both upper and lower indices and has the only non-vanishing components
\be
Y^{\alpha s}{}_{\beta s} = \delta^\alpha_\beta \, .
\label{eq:Y}
\ee
There is also a universal weight term, $+ \omega \partial_N \Lambda^N V^M$. The constant $\omega$ depends on the number $n=11-d$ of external dimensions as $\omega = - \frac{1}{n-2}$ and for us $\omega = - 1/7$. The gauge parameters themselves are chosen to have specific weight $\lambda_\Lambda = 1/7$, which cancels that arising from the $\omega$ term. 

In conventional geometry, diffeomorphisms are generated by the Lie derivative and form a closed algebra under the Lie bracket. 
The algebra of generalised diffeomorphisms involves the $E$-bracket,
\be
[ U,V]_E = \frac{1}{2} \left( \mathcal{L}_U V - \mathcal{L}_V U \right) \,.
\ee
The condition for closure of the algebra is
\be
\mathcal{L}_U \mathcal{L}_V - 
\mathcal{L}_V \mathcal{L}_U 
= \mathcal{L}_{[U,V]_E}  
\ee
which does not happen automatically. A universal feature in all exceptional field theories is that we need to impose the so called {\it{section condition}} \cite{Hull:2009zb,Berman:2011cg,Coimbra:2011ky,Berman:2012vc} so the algebra closes. This is the following constraint determined by the Y-tensor
\be
Y^{MN}{}_{PQ} \partial_M \partial_N = 0 \,,
\ee
which implies the form \eqref{eq:constraint} given before.

From the definition of the generalised Lie derivative \eqref{eq:gld} and the Y-tensor \eqref{eq:Y}, we can write down the transformation rules for the components $V^\alpha$ and $V^s$, which are 
\begin{equation}
 \begin{split}
  \gL_\Lambda V^\alpha &= \Lambda^M \partial_M V^\alpha - V^\beta \partial_\beta \Lambda^\alpha - \frac17 V^\alpha \partial_\beta \Lambda^\beta + \frac67 V^\alpha \partial_\beta \Lambda^\beta \,, \\
  \gL_\Lambda V^s &= \Lambda^M \partial_M V^s + \frac67 V^s \partial_\beta \Lambda^\beta - \frac87 V^s \partial_s \Lambda^s \,.
 \end{split}
\label{eq:gldcpts} 
\end{equation}
Then by requiring the Leibniz property for the generalised Lie derivative, we can derive the transformation rules for tensors in other representations of $\G$, such as the generalised metric $\gM_{MN}$. (The form fields must be treated separately, see Appendix \ref{sec:tensorhier}.)

In doing so, we also need to specify the weight $\lambda$ of each object. It is conventional to choose the generalised metric to have weight zero under generalised diffeomorphisms. Meanwhile, the sequence of form fields $\Aa, \Ab, \Ac,\dots$ are chosen to have weights $\lambda_\Aa = 1/7$, $\lambda_{\Ab} = 2/7$, $\lambda_{\Ac} = 3/7$ and so on. Finally, we take the external metric $g_{\mu \nu}$ to be a scalar of weight $2/7$. 

In the above we have only treated the infinitesimal, local $\G$ symmetry. This should be related to finite $\G$ transformations by exponentiation. The relation between the exponentiated generalised Lie derivative and the finite transformations are quite nontrivial due to the presence of the section condition. For double field theory there are now are series of works dealing with this issue \cite{Hohm:2012gk,Park:2013mpa, Berman:2014jba,Hull:2014mxa,Naseer:2015tia,Rey:2015mba} and recently the EFT case has been studied in \cite{Chaemjumrus:2015vap,Rey:2015mba}.

The other diffeomorphism symmetry of the action consists of external diffeomorphisms, parametrised by vectors $\xi^\mu$. These are given by the usual Lie derivative
\be
\delta_\xi V^\mu \equiv L_\xi V^\mu 
= \xi^\nu D_\nu V^\mu - V^\nu D_\nu \xi^\mu 
+ \hat\lambda_V D_\nu \xi^\nu V^\mu \,,
\ee
with partial derivatives replaced by the derivative $D_\mu$ which is covariant under internal diffeomorphisms, and explicitly defined by
\be
D_\mu = \partial_\mu - \delta_{\Aa_\mu} \,.
\label{eq:Dmu} 
\ee
The weight $\hat\lambda_V$ of a vector under external diffeomorphisms is independent of that under generalised diffeomorphisms.

For this to work, the gauge vector $\Aa_\mu$ must transform under generalised diffeomorphisms as
\be
\delta_\Lambda \Aa_\mu{}^M= D_\mu \Lambda^M\,.
\label{eq:deltaLambdaAa} 
\ee
The external metrics and form fields then transform under external diffeomorphisms in the usual manner given by the Leibniz rule, while the generalised metric is taken to be a scalar, $\delta_\xi \gM_{MN} = \xi^\mu D_\mu \gM_{MN}$. 

In addition, we need to consider the gauge transformations of the remaining form fields \eqref{eq:forms}. Deriving the correct gauge transformations and field strengths is a non-trivial exercise, and we defer the presentation to Appendix \ref{sec:tensorhier}. 

Each individual term in the general form of the action \eqref{eq:S} is separately invariant under generalised diffeomorphisms and gauge transformations. The external diffeomorphisms though mix the various terms and so by requiring invariance under these transformations one may then fix the coefficients of the action. 

An alternative derivation of the generalised Lie derivative is the following. 
We consider a general ansatz for the generalised Lie derivative acting on elements in the 2-dimensional and singlet representation
\begin{equation}
 \begin{split}
  \gL_\Lambda V^\alpha &= \Lambda^M \partial_M V^\alpha - V^\beta \partial_\beta \Lambda^\alpha + a V^\alpha \partial_\beta \Lambda^\beta + b V^\alpha \partial_\beta \Lambda^\beta \,, \\
  \gL_\Lambda V^s &= \Lambda^M \partial_M V^s + c V^s \partial_\beta \Lambda^\beta + d V^s \partial_s \Lambda^s \,.
 \end{split}
\end{equation}
We can fix the coefficients $a$, $b$, $c$, $d$ as follows. We require a singlet $\Delta_s$ and $\epsilon_{\alpha\beta}$ to define an invariant, i.e.
\begin{equation}
 \gL_\Lambda \left(\epsilon_{\alpha\beta} \Delta_s\right) = 0 \,.
\end{equation}
This property allows us to define the unit-determinant generalised metric \eqref{eq:unitdetgenmetric}. Furthermore, we require that the algebra of generalised Lie derivatives closes subject to a section condition. Requiring this to allow for two inequivalent solutions then fixes the coefficients $a$, $b$, $c$, $d$ above and -- up to a redefinition -- reproduces \eqref{eq:gldcpts}. This definition of the generalised Lie derivative fits in the usual pattern of generalised diffeomorphism in EFT described by the Y-tensor \cite{Berman:2012vc} deformation of the Lie derivative.

\section{Relationship to Supergravity}
\label{sec:supergravity} 

Our $\G$ exceptional field theory is equivalent to 11-dimensional and 10-dimensional IIB supergravity, in a particular splitting inspired by Kaluza-Klein reductions. In this section, we present the details of this split and give the precise relationships between the fields of the exceptional field theory and those of supergravity.\footnote{In general, our actions will have the same relative normalisations as that in the book \cite{Ortin:2004ms} although we use the opposite signature. The procedure that we use is essentially the same as carried out in the exceptional field theory literature, see for instance \cite{Hohm:2013vpa, Baguet:2015xha} for detailed descriptions of the M-theory and IIB cases.}

\subsection{Metric terms} 

Let us consider first the $(n+d)$-dimensional Einstein-Hilbert term
\be
S = \int \dd^{n+d}x \sqrt{G} R \,.
\ee
This discussion applies equally to 11-dimensional supergravity and type IIB supergravity. In both cases, $n$ is the number of ``external'' dimensions and $d$ the number of internal. So $n=9$ always but $d=2$ in the 11-dimensional case and $d=1$ in the IIB case. 

The $(n+d)$-dimensional coordinates $x^{\hmu}$ are split into $x^\mu$, $\mu = 1,\dots,9$ and $y^m$, $m=1,\dots,d$. After splitting the $(n+d)$-dimensional flat coordinates $\ha$ into $n$-dimensional flat coordinates $a$ and $d$-dimensional flat coordinates $\bm$, we write the $(n+d)$-dimensional vielbein as
\be
E^{\ha}{}_{\hmu} = \begin{pmatrix}
 \phi^{\omega/2} e^a{}_\mu & 0 \\
 A_\mu{}^m \phi^{\bm}{}_m & 
 \phi^{\bm}{}_m 
\end{pmatrix} \,.
\ee
Fixing this form of the vielbein breaks the $\mathrm{SO}(1,n+d-1)$ Lorentz symmetry to $\mathrm{SO}(1,n-1) \times \mathrm{SO}(d)$. Note however that we continue to allow the fields of the theory to depend on all the coordinates, so at no point do we carry out a dimensional reduction.

Here we treat $e^a{}_\mu$ as the vielbein for the external metric $g_{\mu \nu}$ and can think of $\phi^{\bm}{}_m$ as the vielbein for the internal metric $\phi_{mn}$. Here $\phi \equiv \det \phi_{mn}$.
The corresponding form of the metric is
\be
G_{\hmu \hnu} = 
\begin{pmatrix} \phi^{\omega} g_{\mu \nu} + A_\mu{}^p A_\nu{}^q \phi_{pq} & A_\mu{}^p \phi_{pn} \\
A_\nu{}^p \phi_{pm} & \phi_{mn} 
\end{pmatrix} \,.
\label{eq:GKK}
\ee
The constant $\omega$ is fixed in order to obtain the Einstein-Hilbert term for the metric $g_{\mu \nu}$ and is the same as in Section \ref{sec:theory}: $\omega=- \frac{1}{n-2}=-1/7$.

Diffeomorphisms $\xi^{\hmu}$ split into internal $\Lambda^m$ and external $\xi^\mu$ transformations. We define covariant derivatives $D_\mu = \partial_\mu  - \delta_{A_\mu}$ which are covariant with respect to internal diffeomorphisms
\begin{equation}
\begin{aligned}
D_\mu e^a{}_\nu &= \partial_\mu e^a{}_\nu - A_\mu{}^m \partial_m e^a{}_\nu + \omega \partial_n A_\mu{}^n e^a{}_\nu \,, \\
D_\mu \phi^{\bm}{}_m &= \partial_\mu \phi^{\bm}{}_m - A_\mu{}^n \partial_n \phi^{\bm}{}_m - \partial_m A_\mu{}^n \phi^{\bm}{}_n \,.
\end{aligned}
\end{equation}
The use of the letter $D_\mu$ here and also for the covariant derivatives \eqref{eq:Dmu} appearing in the EFT is no accident. Indeed, on solving the section condition, those in EFT reduce exactly to the expressions here.

One can think of $e^a{}_\mu$ as carrying density weight $-\omega$ under internal diffeomorphisms. The Kaluza-Klein vector appears as a connection for internal diffeomorphisms, the field strength is 
\be
F_{\mu \nu}{}^m = 2 \partial_{[\mu} A_{\nu]}{}^m - 2 A_{[\mu|}{}^n \partial_n A_{\nu]}^m \,.
\ee 
To obtain external diffeomorphisms one must add a compensating Lorentz transformation of the vielbein. The resulting expressions are not covariant with respect to internal diffeomorphisms, but can be improved by adding a field dependent internal transformation (with parameter $-\xi^\nu A_\nu{}^m$) to each transformation rule. This leads to the definition of external diffeomorphisms in our split theory
\begin{equation}
\begin{aligned}
\delta_\xi e^a{}_\mu &= \xi^\nu D_\nu e^a{}_\mu + D_\mu \xi^\nu e^a{}_\nu \,, \\
\delta_\xi \phi^{\bm}{}_m &= \xi^\nu D_\nu \phi^{\bm}{}_m \,, \\
\delta_\xi A_\mu{}^m &= \xi^\nu F_{\nu \mu}{}^m + \phi^{\omega} \phi^{mn} g_{\mu \nu} \partial_n \xi^\mu \,.
\end{aligned}
\end{equation}
It is convenient to also define a derivative 
\be
D_m e^a{}_\mu = \partial_m e^a{}_\mu + \frac{\omega}{2} e^a{}_\mu  \partial_m \ln \phi \,.
\ee
One can then show that the Einstein-Hilbert term $S = \int \dd^{n+d}x\sqrt{G} R$ becomes
\be
\begin{split}
\int \dd^n x \dd^d y \, \sqrt{|g|} 
\Bigg[ & 
\hat{R} - \frac{1}{4} \phi^{-\omega} F^{\mu \nu m} F_{\mu \nu m} 
 \\ 
& 
+ \frac{1}{4} g^{\mu \nu} D_\mu \phi^{mn} D_\nu \phi_{mn} 
+ \frac{1}{4} \omega g^{\mu \nu} D_\mu \ln \phi D_\nu \ln \phi 
\\ &
-  \omega \left( D_\mu D^\mu \ln \phi + \frac{1}{2} D_\mu \ln g  D^\mu \ln \phi \right) 
\\ 
& 
+ \phi^{\omega} \left(
R_{int} ( \phi) + \frac{1}{4} \phi^{mn} D_m g^{\mu \nu} D_n g_{\mu \nu} 
+ \frac{1}{4} \phi^{mn} D_m \ln g D_n \ln g 
\right)\Bigg] 
 \\
 - \int \dd^n x \dd^d y \partial_m & \left( \sqrt{|g|}  \phi^{\omega} \phi^{mn} D_n \ln g \right)  \,. 
\end{split} 
\label{eq:EHresult}
\ee
Here $R_{int}(\phi)$ is the object given by the usual formula for the Ricci scalar applied to the internal metric $\phi_{mn}$, using only the $\partial_m$ derivatives (and not involving any determinant $e$ factors). Note that this vanishes when $d=1$ i.e. for the IIB splitting. Meanwhile, $\hat{R}$ is the improved Ricci scalar in which all derivatives that appear are $D_\mu$ rather than $\partial_\mu$.

\subsection{Splitting of 11-dimensional supergravity}

Now, we come to 11-dimensional supergravity. We write the (bosonic) action as
\be
S_{11} = \int d^{11} x \sqrt{G}\left( 
R 
- \frac{1}{48} \hat F^{\hmu \hnu \hrho \hlambda} 
\hat F_{\hmu \hnu \hrho \hlambda} 
+ \frac{1}{(144)^2} \frac{1}{\sqrt{G}} \varepsilon^{\hmu_1 \dots \hmu_{11}} 
\hat F_{\hmu_1 \dots \hmu_4} 
\hat F_{\hmu_5 \dots \hmu_8} 
\hat C_{\hmu_9 \dots \hmu_{11}} 
\right) \,.
\ee
We will slightly adapt our notation here. The index $\alpha = 1,2$ is used to denote internal indices, while we denote the internal components of the metric by $\gamma_{\alpha \beta}$, i.e. with respect to the previous section $\phi_{mn} \rightarrow \gamma_{\alpha \beta}$. 

The four-form field strength is as usual
\be
\hat F_{\hmu \hnu \hrho \hlambda} = 4 \partial_{[\hmu} \hat C_{\hnu \hrho \hlambda]} \,.
\ee
The degrees of freedom arising from the metric are then the external metric $g_{\mu \nu}$, the Kaluza-Klein vector $A_\mu{}^\alpha$ with field strength
\be
F_{\mu \nu}{}^\alpha = 2 \partial_{[\mu} A_{\nu]}{}^\alpha - 2 A_{[\mu|}{}^\beta \partial_\beta A_{\nu]}^\alpha \,,
\ee
and the internal metric $\gamma_{\alpha \beta}$. 

The three-form field gives $n$-dimensional forms $\hat C_{\mu \nu \rho}$, $\hat C_{\mu \nu \alpha}$ and $\hat C_{\mu \alpha \beta}$. In order to obtain fields which have better transformation properties under the symmetries in the split (both diffeomorphisms and gauge transformations), one redefines the form field components by flattening indices with $E_{\ha}{}^{\hmu}$ and then curving them with $E^{\ha}{}_\mu$, so that for instance $\bA_{\mu \alpha \beta} \equiv E^{\ha}{}_\mu E_{\ha}{}^{\hmu} \hat{C}_{\hmu \alpha \beta}$,
\be
\begin{split}
\bA_{\mu \alpha \beta } & = \hat C_{\mu \alpha \beta } \,,\\ 
\bB_{\mu \nu \alpha} & = \hat C_{\mu \nu \alpha} - 2 A_{[\mu}{}^\beta \hat C_{\nu] \alpha \beta } \,,\\ 
\bC_{\mu \nu \rho } & = \hat C_{\mu\nu \rho} 
 - 3 A_{[\mu}{}^\alpha \hat C_{\nu \rho] \alpha} 
+ 3 A_{[\mu}{}^\alpha A_{\nu}{}^\beta \hat C_{\rho] \alpha \beta  } \,.
\label{eq:11dABC}
\end{split}
\ee
The fields defined in this way are such that they transform according to their index structure under internal diffeomorphisms $\Lambda^m$ acting as the Lie derivative. 
The field strengths may be similarly redefined
\be
\begin{split}
\bF_{\mu \nu \alpha \beta } & = 2 D_{[\mu } \bA_{\nu ] \alpha \beta} +2 \partial_{[\alpha} \bB_{|\mu \nu| \beta ]}\,,\\
\bF_{\mu \nu \rho \alpha} & = 3 D_{[\mu} \bB_{\nu \rho] \alpha} + 3 F_{[\mu \nu}{}^\beta \bA_{\rho] \alpha \beta } 
- \partial_\alpha \bC_{\mu \nu \rho} \,,\\
\bF_{\mu \nu \rho \sigma} & = 4 D_{[\mu } \bC_{\nu \rho \sigma]} 
+ 6 F_{[\mu \nu}{}^\alpha \bB_{\rho \sigma] \alpha } \,.
\end{split} 
\ee
Here we have the covariant derivative $D_\mu = \partial_\mu - \delta_{A_\mu}$ introduced above. 

The kinetic terms for the gauge fields may be easily decomposed by going to flat indices and then using the above redefinitions. 
Including the Kaluza-Klein vector, one finds the total gauge kinetic terms
\be
- \frac{1}{4} \gamma^{1/7} \gamma_{\alpha \beta} F_{\mu \nu}{}^{ \alpha} F^{\mu \nu \beta}
- \frac{1}{8} \gamma^{1/7} \gamma^{\alpha \beta} \gamma^{\gamma \delta} \bF_{\mu \nu \alpha \gamma} \bF^{\mu \nu}{}_{\beta \delta} 
- \frac{1}{12} \gamma^{2/7} \gamma^{\alpha \beta} \bF_{\mu \nu \rho}{}_{\alpha} \bF_{\mu \nu \rho \beta} 
-\frac{1}{48} \gamma^{3/7} \bF^{\mu \nu \rho \sigma} \bF_{\mu \nu \rho \sigma} 
\label{eq:11dgkin}
\ee
Finally, consider the Chern-Simons term. A very convenient way of treating this reduction is to use the trick of rewriting the Chern-Simons term as a manifestly gauge invariant term in one dimension higher (note this fictitious extra dimension has nothing to do with the extra coordinate introduced in EFT). Thus one writes
\be
S_{CS} = -  
\frac{1}{4\cdot (144)^2} 
\int \dd^{12}x \varepsilon^{ \hmu_1 \dots \hmu_{12} } 
\hat F_{\hmu_1 \dots \hmu_4} 
\hat F_{\hmu_5 \dots \hmu_8} 
\hat F_{\hmu_9 \dots \hmu_{12}} \,.
\label{eq:MCSlift}  
\ee
The variation of this is
\be
\frac{3}{(144)^2}
 \int \dd^{12}x
\partial_{\hmu_{12}} 
\left(
\varepsilon^{ \hmu_1 \dots \hmu_{12} } 
\hat F_{\hmu_1 \dots \hmu_4} 
\hat F_{\hmu_5 \dots \hmu_8} 
\delta \hat C_{\hmu_9 \dots \hmu_{11}}
\right) \,,
\ee
which is a total derivative as expected. This term can be decomposed according to the above splitting. One obtains
\be
S_{CS} = 
- 
\frac{1}{8 \cdot 12 \cdot 144} 
\int \dd^{10}x \dd^2 y \varepsilon^{\mu_1 \dots \mu_{10}} \varepsilon^{\alpha \beta} \left(
3 \bF_{\alpha \beta \mu_1 \mu_2} 
\bF_{\mu_3 \dots \mu_6} 
\bF_{\mu_7 \dots \mu_{10}} 
- 8
\bF_{\alpha \mu_1 \mu_2\mu_3} 
\bF_{\beta \mu_4 \mu_5\mu_6} 
\bF_{\mu_7 \dots \mu_{10}} 
\right) \,.
\label{eq:MCStermsplit} 
\ee

\subsection{The EFT/M-theory dictionary}
\label{sec:reductionM}

Now, we take our $\G$ EFT described in section \ref{sec:theory} and impose the M-theory section condition, $\partial_s = 0$. Thus, the fields of our theory depend on the coordinates $x^\mu$ and $y^\alpha$, which are taken to be the coordinates of 11-dimensional supergravity in the $9+2$ splitting described above. 

The metric-like degrees of freedom are easily identified. The external metric $g_{\mu \nu}$ used in the $\G$ EFT is simply that appearing in the splitting of 11-dimensional supergravity. Meanwhile, the generalised Lie derivative tells us how the generalised metric should be parametrised in spacetime, by interpreting the transformation rules it gives in the M-theory section in terms of internal diffeomorphisms. For instance, one sees from \eqref{eq:gldcpts} that with $\partial_s = 0$
\be
\begin{split} 
\delta_\Lambda \gM_{\alpha \beta} & = \Lambda^\gamma \partial_\gamma \gM_{\alpha \beta} + \partial_\alpha \Lambda^\gamma \gM_{\gamma \beta} 
+ \partial_\beta \Lambda^\gamma \gM_{\alpha \gamma} 
+ \frac{2}{7} \partial_\gamma \Lambda^\gamma \gM_{\alpha \beta} \,,\\ 
\delta_\Lambda \gM_{ss} & = \Lambda^\gamma \partial_\gamma \gM_{ss} - \frac{12}{7} \partial_\gamma \Lambda^\gamma \gM_{ss} \,.
\end{split} 
\ee
which tells us that $\MM_{\alpha \beta}$ transforms as a rank two tensor of weight $2/7$ under internal diffeomorphisms while $\MM_{ss}$ is a scalar of weight $-12/7$, so that:
\begin{align}
\MM_{\alpha\beta} &= \gamma^{1/7}\gamma_{\alpha\beta} \, , &
\MM_{ss} &= \gamma^{-6/7} \, .
\label{eq:genmetricM}
\end{align}
It is straightforward to check that after inserting this into the EFT action that one obtains the correct action resulting from the 11-dimensional Einstein-Hilbert term, given by \eqref{eq:EHresult}. This verification involves the scalar potential, the scalar kinetic terms and the external Ricci scalar. The part of this calculation involving the external derivatives $D_\mu$ works almost automatically after identifying the Kaluza-Klein vector $A_\mu{}^\alpha$ with the $\alpha$ component of the vector $A_\mu{}^M$, such that the derivatives $D_\mu$ used in the EFT become those defined above in the splitting of supergravity. 

Now we come to the form fields of EFT. The ones that appear with kinetic terms in the action are $A_\mu{}^M$, $B_{\mu \nu}{}^{\alpha \beta s}$, $C_{\mu \nu \rho}{}^{\alpha \beta s s}$. These are related to the Kaluza-Klein vector $A_\mu{}^\alpha$ and the redefined fields \eqref{eq:11dABC} by the following 
\be
\begin{split} 
A_\mu{}^s & \equiv  \frac{1}{2} \epsilon^{\alpha \beta} \bA_{\mu \alpha \beta} 
\\ 
B_{\mu \nu}{}^{\alpha s} & \equiv  \epsilon^{\alpha \beta}  \bB_{\mu \nu \beta} -  \frac{1}{2} \epsilon^{\gamma \delta} A_{[\mu}{}^\alpha \bA_{\nu] \gamma \delta} 
\\ 
C_{\mu \nu \rho}{}^{\alpha \beta s} & \equiv  \epsilon^{\alpha \beta} \bC_{\mu \nu \rho} 
- 2 \frac{1}{2} \epsilon^{\gamma \delta} A_{[\mu}{}^\alpha A_\nu{}^\beta \bA_{\rho] \gamma \delta}
\end{split}
\label{eq:EFTABC}
\ee
These definitions are such that the field strength components are
\be
\begin{split} 
\Fa_{\mu \nu}{}^\alpha & = F_{\mu \nu}{}^\alpha \,,\\ 
\Fa_{\mu \nu}{}^s & =  \frac{1}{2}  \epsilon^{\alpha \beta}  \bF_{\mu \nu \alpha \beta} \,, \\
\Fb_{\mu \nu \rho}{}^{\alpha s} & =  \epsilon^{\alpha \beta} \bF_{\mu \nu \rho \beta} \,,\\ 
\Fc_{\mu \nu \rho \sigma}{}^{\alpha \beta s} & =  \epsilon^{\alpha \beta} \bF_{\mu \nu \rho \sigma} \,.
\end{split} 
\ee
We can then straightforwardly write down the gauge kinetic terms of the EFT action, which in this section and with the parametrisation \eqref{eq:genmetricM} of the generalised metric are given by
\be
\begin{split}
& - \frac{1}{4} \gamma^{1/7} \gamma_{\alpha \beta} \Fa_{\mu \nu}{}^\alpha \Fa^{\mu \nu \beta} 
- \frac{1}{4} \gamma^{1/7} \gamma^{-1} \Fa_{\mu \nu}{}^s \Fa^{\mu \nu s} 
\\ & - \frac{1}{12} \gamma^{2/7} \gamma^{-1} \gamma_{\alpha \beta} \Fb_{\mu \nu \rho}{}^{\alpha s} \Fb^{\mu \nu \rho \beta s} 
- \frac{1}{4 \cdot 24} \gamma^{3/7} \gamma^{-1} \gamma_{\alpha \gamma} \gamma_{\beta \delta} \Fc_{\mu \nu \rho \sigma}{}^{[\alpha \beta ] s} 
\Fc^{\mu \nu \rho \sigma [ \gamma \delta ] s} \,,
\end{split} 
\ee
and show that this automatically reduces to those of 11-dimensional supergravity, \eqref{eq:11dgkin}. 

Similarly, one can show that we obtain the correct Chern-Simons term \eqref{eq:MCStermsplit}. The remaining gauge fields that appear in EFT, which are a four-form, five-form and six-form are not dynamical. The action must always be complemented by a self-duality relation that relates $p$-form field strengths to their magnetic duals. This is a natural consequence of the formalism where we have included both electric and magnetic descriptions in the action simultaneously.

\subsection{Splitting of 10-dimensional type IIB supergravity} 

The (bosonic) (pseudo-)action of type IIB can be written as
\be
\begin{split}
S & = 
\int \dd^{10} x \sqrt{G} \Big(
R 
+ \frac{1}{4} G^{\hmu \hnu} \partial_{\hmu} \cH_{\alpha \beta} \partial_{\hmu} \cH^{\alpha \beta} 
- \frac{1}{12} \cH_{\alpha \beta} \hat F_{\hmu \hnu \rho}{}^{\alpha} \hat F^{\hmu \hnu \rho \beta} 
- \frac{1}{480} \hat F_{\hmu_1 \dots \hmu_5}\hat F^{\hmu_1 \dots \hmu_5} 
\Big) 
\\ & 
+ \frac{1}{24 \cdot 144} \int \dd^{10}x \epsilon_{\alpha \beta} \epsilon^{\hmu_1 \dots \hmu_{10}} \hat C_{\hmu_1 \hmu_2 \hmu_3 \hmu_4} 
\hat F_{\hmu_5 \hmu_6\hmu_7}{}^\alpha \hat F_{\hmu_8 \hmu_9 \hmu_{10}}{}^\beta \,.
\end{split} 
\ee
This action must be accompanied by the duality relation for the self-dual five-form
\be
\hat{F}_{\hmu_1 \dots \hmu_5} = \frac{1}{5!} \sqrt{G} \epsilon_{\hmu_1 \dots \hmu_{10}} \hat F^{\hmu_6 \dots \hmu_{10} }
\ee
The field strengths themselves are written as
\be
\hat F_{\hmu \hnu \rho}{}^\alpha = 3 \partial_{[\hmu} \hat C_{\hnu \rho]}{}^\alpha
\ee
and\be
\hat F_{\hmu_1 \dots \hmu_5} = 5 \partial_{[\hmu_1} \hat C_{\hmu_2 \dots \hmu_5]} 
+ 5  \epsilon_{\alpha \beta} \hat C_{[\hmu_1 \hmu_2}{}^\alpha \hat F_{\hmu_3 \hmu_4 \hmu_5]}{}^{\beta} 
\ee
We carry out a $9+1$ split of the coordinates. In this section, we will denote the internal index by $s$ (for singlet), and the single internal metric component by $\phi_{ss} \equiv \phi$. The metric then gives the external metric, $g_{\mu\nu}$, the Kaluza-Klein vector $A_\mu{}^s$ with field strength
\be
F_{\mu \nu}{}^s = 2 \partial_{[\mu} A_{\nu]}{}^s - 2 A_{[\mu}{}^s \partial_s A_{\nu]}{}^s \,.
\ee
The scalars $\cH_{\alpha \beta}$ are trivially reduced using the decomposition of the metric to give
\be
+ \frac{1}{4} g^{\mu \nu} D_\mu \cH_{\alpha \beta} D_\nu \cH^{\alpha \beta} 
+ \frac{1}{4} \phi^{-8/7} \partial_s \cH_{\alpha \beta} \partial_s \cH^{\alpha \beta}\,.
\label{eq:Skinred}
\ee
From the two-form, using the standard trick involving contracting with $E^{\ha}{}_\mu E_{\ha}{}^{\hmu}$ to obtain appropriate decompositions of the forms, we find the components
\be
\begin{split}
\bC_{\mu s}{}^{\alpha} & \equiv \hat C_{\mu s}{}^\alpha \,,\\
\bC_{\mu \nu}{}^{\alpha}  & \equiv \hat C_{\mu \nu}{}^\alpha + 2 A_{[\mu}{}^s \hat C_{\nu] s}{}^\alpha \,,
\end{split} 
\ee
so that the field strengths are
\be
\begin{split}
\bar{F}_{\mu \nu s}{}^\alpha & \equiv \hat{F}_{\mu \nu s}{}^\alpha  \\
& = 2 D_{[\mu} \bC_{\nu] s}{}^\alpha + \partial_s \bC_{\mu \nu}{}^\alpha \,,  \\
\bar{F}_{\mu \nu \rho}{}^\alpha & \equiv \hat F_{\mu \nu \rho}{}^\alpha - 3 A_{[\mu}{}^s \hat{F}_{\nu \rho]s} \\
 & = 3 D_{[\mu} \bC_{\nu \rho]}{}^\alpha -  3 F_{[\mu \nu}{}^s \bC_{\rho]s}{}^\alpha \,.
\end{split} 
\ee
Note that the two-form doublet $\hat C_{\mu \nu}{}^\alpha$ consists of the NSNS-two-form $B_{\mu \nu}$ as its first component and the RR-two-form $C_{\mu \nu}$ as its second component. For the four-form one has similarly 
\be
\begin{split}
\bC_{\mu \nu \rho s} & \equiv \hat C_{\mu \nu \rho s}\,, \\
\bC_{\mu \nu \rho \sigma} & \equiv \hat C_{\mu \nu \rho \sigma}  + 4 A_{[\mu}{}^s \hat C_{\nu \rho \sigma] s} \,,
\end{split} 
\ee
with field strengths
\be
\begin{split}
\bF_{\mu \nu \rho \sigma s}  & \equiv \hat F_{\mu \nu \rho \sigma s}  \\
 & = 4 D_{[\mu} \bC_{\nu \rho \sigma]s} + \partial_s \bC_{\mu \nu \rho \sigma} 
+ \epsilon_{\alpha \beta} \left( 2 \bC_{s[\mu}{}^\alpha \bF_{\nu \rho \sigma]}{}^\beta
+ 3 \bC_{[\mu \nu} \bF_{\rho \sigma]z}{}^\beta \right) \,, \\
\bF_{\mu \nu \rho \sigma \lambda}  & \equiv \hat F_{\mu \nu \rho \sigma \lambda }  
-5 A_{[\mu}{}^s \hat F_{\nu \rho \sigma \lambda]s} 
 \\
 & = 5 D_{[\mu} \bC_{\nu \rho \sigma \lambda]} -20 F_{[\mu \nu}{}^s \bC_{\rho \sigma \lambda]s}
+ 5 \epsilon_{\alpha \beta} 
\bC_{[\mu \nu} \bF_{\rho \sigma \lambda]}{}^\beta \,.
\end{split} 
\ee 
In terms of these objects, the duality relation becomes 
\be
\bF_{\mu \nu \rho \sigma s} = \frac{1}{5!} \phi^{4/7} g^{1/2} \epsilon_{\mu \nu \rho \sigma }{}^{\nu_1 \dots \nu_5} 
\bF_{\nu_1 \dots \nu_5} \,.
\label{eq:IIBstarF}
\ee
These definitions lead to the following kinetic terms in the Lagrangian
\be
\begin{split}  & 
- \frac{1}{4} \phi^{8/7} F_{\mu \nu}{}^s F^{\mu \nu s} 
- \frac{1}{4} \cH_{\alpha \beta} \phi^{-6/7} \bF_{\mu \nu s}{}^\alpha \bF^{\mu \nu}{}_s{}^\beta
- \frac{1}{12} \cH_{\alpha \beta} \phi^{2/7} \bF_{\mu \nu \rho}{}^{ \alpha} \bF^{\mu \nu \rho \beta} \\
& 
- \frac{1}{96} \phi^{4/7} \bF_{\mu \nu \rho \sigma s} \bF^{\mu \nu \rho \sigma}{}_s 
- \frac{1}{480} \phi^{4/7} \bF_{\mu_1 \dots \mu_5} \bF^{\mu_1 \dots \mu_5} \,.
\end{split} 
\label{eq:IIBFormAction}
\ee
Finally, consider the Chern-Simons term which can be written in one dimension higher as 
\be
\begin{split} 
- \frac{1}{5 \cdot 24 \cdot 144} 
& \int \dd^{11} x 
\epsilon_{\alpha \beta} \epsilon^{\hmu_1 \dots \hmu_{11}}
\hat F_{\hmu_1 \hmu_2\hmu_3}{}^\alpha 
\hat F_{\hmu_4 \hmu_5 \hmu_{6}}{}^\beta 5
\partial_{\hmu_{11}} \hat C_{\hmu_7 \hmu_8 \hmu_9 \hmu_{10} } 
=\\ & 
- \frac{1}{5\cdot 24 \cdot 144} 
\int \dd^{11} x 
\epsilon_{\alpha \beta} \epsilon^{\hmu_1 \dots \hmu_{11}}
\hat F_{\hmu_1 \hmu_2\hmu_3}{}^\alpha 
\hat F_{\hmu_4 \hmu_5 \hmu_{6}}{}^\beta 
\hat F_{\hmu_7 \hmu_8 \hmu_9 \hmu_{11} } \,.
\end{split}
\ee
Under the split, this becomes
\be
- \frac{1}{5\cdot 24 \cdot 144} 
\int \dd^{10} x \dd y 
\epsilon_{\alpha \beta} \epsilon^{\mu_1 \dots \mu_{10}}
\left(
6 \bF_{\mu_1 \mu_2 s}{}^\alpha 
\bF_{\mu_3 \mu_4 \mu_{5}}{}^\beta 
\bF_{\mu_6 \dots  \mu_{10} }
+ 5
 \bF_{\mu_1 \mu_2 \mu_3}{}^\alpha 
\bF_{\mu_4 \mu_5 \mu_{6}}{}^\beta 
\bF_{ \mu_7 \dots\mu_{10} s }
\right)
\label{eq:IIBCSsplit} 
 \,.
\ee

\subsection{The EFT/Type IIB dictionary} 
\label{sec:reductionF}

Now, we take our $\G$ EFT and impose the IIB section, $\partial_\alpha = 0$. The fields then depend on the coordinates $x^\mu$ and $y^s$, which become the coordinates of 10-dimensional type IIB supergravity in the $9+1$ split we have described above. 

The external metric can be immediately identified. The components of the generalised metric are 
\begin{align}
\MM_{\alpha\beta} &= \phi^{-6/7} \cH_{\alpha\beta} \, ,  &
\MM_{ss} &= \phi^{8/7} \, .
\label{eq:genmetricIIB}
\end{align}
The Kaluza-Klein vector $A_\mu{}^s$ can be identified as the $s$ component of the gauge field $A_\mu{}^M$. One can then verify the reduction of the scalar potential, scalar kinetic terms and external Ricci scalar, and verify that they give the expected reduction of the Einstein-Hilbert term, \eqref{eq:EHresult} and scalar kinetic terms \eqref{eq:Skinred}. For completeness let us give here the parametrisation of $\cH_{\alpha \beta}$ in terms of $\tau = C_0 + i e^{-\varphi}$: 
\begin{equation}
\cH_{\alpha\beta} = \frac{1}{\tau_2}
\begin{pmatrix}
1 & \tau_1 \\ \tau_1 & |\tau|^2
\end{pmatrix} = e^\varphi
\begin{pmatrix}
1 & C_0 \\ C_0 & C_0^2 + e^{-2\varphi}
\end{pmatrix} \, .
\end{equation}
For the remaining forms, we need the following definitions
\be
\begin{split}
A_\mu{}^\alpha & \equiv \bC_{\mu s}{}^\alpha \,,
\\ 
B_{\mu \nu}{}^{\alpha s} & \equiv  \bC_{\mu \nu}{}^{\alpha} - A_{[\mu}{}^s \bC_{\nu] s}{}^\alpha 
  \,, \\
C_{\mu \nu \rho}{}^{\alpha \beta s} 
& \equiv \epsilon^{\alpha \beta} \bC_{\mu \nu \rho}
+3 \bC_{[\mu|s|}{}^{[\alpha} \bC_{\nu \rho]}{}^{\beta]} 
+ 4 \bC_{[\mu|s|}{}^\alpha \bC_{\nu|s|}{}^\beta A_{\rho]}{}^s \\ 
D_{\mu \nu \rho \sigma}{}^{\alpha \beta ss} & \equiv 
\epsilon^{\alpha \beta} \bC_{\mu \nu \rho \sigma} 
+ 6 \bC_{[\mu \nu}{}^{[\alpha} \bC_{\rho |s|}{}^{\beta]} A_{\sigma]}{}^s \,,
\end{split} 
\ee
such that
\be
\begin{split}
\mathcal{F}_{\mu \nu}{}^{s} & = F_{\mu \nu}{}^s \,, \\ 
\mathcal{F}_{\mu \nu}{}^{\alpha} & = \bF_{\mu \nu s}{}^\alpha \,, \\ 
\Fb_{\mu \nu \rho}{}^{\alpha s} & = \bF_{\mu \nu \rho}{}^\alpha \,, \\ 
\Fc_{\mu \nu \rho \sigma}{}^{\alpha \beta s} & =  \epsilon^{\alpha \beta}  \bF_{\mu \nu \rho \sigma s} \,,\\ 
\Fd_{\mu \nu \rho \sigma \lambda}{}^{\alpha \beta ss} 
& =  \epsilon^{\alpha \beta} \bF_{\mu \nu \rho \sigma \lambda} \,.
\end{split}
\ee
The duality relation that one obtains from the EFT action by varying with respect to $\Delta D_{\mu \nu \rho \sigma}$ is (equation \eqref{eq:Deom}) 
\be
\partial_s \left( \frac{\kappa}{2} \epsilon^{\mu_1 \dots \mu_9} \epsilon_{\alpha \beta} 
\epsilon_{\gamma \delta} \Fd_{\mu_5 \dots \mu_9}{}^{\gamma \delta ss} 
- 2 e \frac{1}{96} \gM_{ss}  \gM_{\alpha \gamma} \gM_{\beta \delta} \Fc^{\mu_1 \dots \mu_4 \gamma \delta s} \right) = 0 
\ee
Here $\kappa$ is the overall coefficient of the topological term \eqref{eq:ToptTerm}. 
After some manipulation one sees that this is consistent with the duality relation imposed in supergravity, given in \eqref{eq:IIBstarF}, if
\be
\kappa = \frac{ 2 }{5!\cdot 96}\,, 
\ee
which is satisfied by our coefficients. 

The kinetic terms of the EFT action are in this parametrisation given by 
\be
\begin{split}  & 
- \frac{1}{4} \phi^{8/7} \Fa_{\mu \nu}{}^s \Fa^{\mu \nu s} 
- \frac{1}{4} \cH_{\alpha \beta} \phi^{-6/7} \Fa_{\mu \nu}{}^\alpha \Fa^{\mu \nu\beta}
- \frac{1}{12} \cH_{\alpha \beta} \phi^{2/7} \Fb_{\mu \nu \rho}{}^{ \alpha s} \Fb^{\mu \nu \rho \beta s} \\
& 
- \frac{1}{96} \phi^{4/7} \cH_{\alpha \gamma} \cH_{\beta \delta} \Fc_{\mu \nu \rho \sigma}{}^{[\alpha \beta]s}  \Fc^{\mu \nu \rho \sigma [ \gamma \delta ] s} \,.
\end{split} 
\ee
Using the above definitions, we find that the first line of this (involving just the Kaluza-Klein vector and the components of the two-form) matches exactly the first line of \eqref{eq:IIBFormAction}. Before discussing the remaining term, we first consider the Chern-Simons term. It is convenient in the IIB section to rewrite this using the Bianchi identities \eqref{eq:Bianchi} given in the appendix. Then the topological term as given in the form \eqref{eq:ToptTermApp} can be written as
\be
\begin{split}
\kappa \int \dd^{10} x d y
& \epsilon^{\mu_1 \dots \mu_{10}} \frac{1}{4} \epsilon_{\alpha \beta} \epsilon_{\gamma \delta} 
\Bigg(
D_{\mu_1} \left( \Fc_{\mu_2 \dots \mu_5}{}^{\alpha \beta s} \Fd_{\mu_6 \dots \mu_{10}}{}^{\gamma \delta ss} \right) 
\\
 & + 4  \Fa_{\mu_1 \mu_2}{}^\alpha  \Fb_{\mu_3 \mu_4 \mu_5}{}^{\beta s}
\Fd_{\mu_6 \dots \mu_{10}}{}^{\gamma \delta ss}
+ \frac{10}{3} \Fb_{\mu_1 \mu_2 \mu_3}{}^{\alpha s} \Fb_{\mu_4 \mu_5 \mu_6}{}^{\beta s} 
 \Fc_{\mu_7 \dots \mu_{10}}{}^{\gamma \delta s} 
\Bigg) \,.
\end{split} 
\ee
The last two lines here give exactly the Chern-Simons term \eqref{eq:IIBCSsplit}. 

The remaining terms give kinetic terms for the field strength. The terms that one obtains differ from those that one obtains from a decomposition of the IIB pseudo-action by a multiplicative factor of $2$, that is the EFT gives 
\be
- \frac{1}{48} \phi^{4/7} \bF_{\mu \nu \rho \sigma s} \bF^{\mu \nu \rho \sigma}{}_s 
- \frac{1}{240} \phi^{4/7} \bF_{\mu_1 \dots \mu_5} \bF^{\mu_1 \dots \mu_5} \,,
\ee
to be compared with the coefficients of $1/96$ and $1/480$ in \eqref{eq:IIBFormAction}. This is as expected due to the use of the self-duality relation which is an equation of motion for the gauge field $D_{\mu \nu \rho \sigma}$. In ten dimensions one has the normalisation $\frac{1}{4} \frac{1}{5!} (F_5)^2$ for the five-form field strength instead of the standard $\frac{1}{2} \frac{1}{5!}$ due to the fact that there are unphysical degrees of freedom which are eliminated by the self-duality relation after varying the IIB action. Here we see only the physical half after using the self-duality relation in the EFT action. The upshot is that strictly speaking the EFT is only equivalent to IIB at the level of the equations of motion, as would be expected given that the IIB action is only a pseudo-action. 

\subsection{Summary} 

The above results display the mapping between the fields of the $\G$ EFT and those of supergravity in a certain Kaluza-Klein-esque split. It is straightforward to relate this back directly to the fields in 11- and 10-dimensions themselves. 

For M-theory, one has for the degrees of freedom coming from the spacetime metric, with the Kaluza-Klein vector $A_\mu{}^\alpha = \gamma^{\alpha \beta} G_{\mu \beta}$,
\begin{equation}
 \begin{split}
\cH_{\alpha\beta} &= \gamma^{-1/2} \gamma_{\alpha\beta} \,, \\
{\cal M}_{ss} &= \gamma^{-6/7} \,, \\
g_{\mu\nu} & = \gamma^{1/7} \left( G_{\mu \nu} - \gamma_{\alpha \beta} A_\mu{}^\alpha A_\nu{}^\beta \right) \,,   \\
  A_\mu{}^s &= \frac12 \epsilon^{\alpha\beta} \hat{C}_{\mu\alpha\beta} \,, \\
  B_{\mu\nu}{}^{\alpha,s} &= \epsilon^{\alpha\beta} \hat{C}_{\mu\nu\beta} + \frac12 \epsilon^{\beta\gamma} A_{[\mu}{}^\alpha \hat{C}_{\nu]\beta\gamma} \,, \\
  C_{\mu\nu\rho}{}^{\alpha\beta,s} &= \epsilon^{\alpha\beta} \left( \hat{C}_{\mu\nu\rho} - 3 A_{[\mu}{}^{\gamma} \hat{C}_{\nu\rho]\gamma} + 2 A_{[\mu}{}^{\gamma} A_{\nu]}{}^{\delta} \hat{C}_{\rho\gamma\delta} \right) \,.
 \end{split} \label{eq:MEFT}
\end{equation}
The inverse relationships, giving the 11-dimensional fields in terms of those in our EFT are
 \begin{equation}
 \begin{split}
  \gamma &= \left( {\cal M}_{ss}\right)^{-7/6} \,, \\
  \gamma_{\alpha\beta} &= \cH_{\alpha\beta} \left( {\cal M}_{ss} \right)^{-7/12} \,, \\
  G_{\mu\nu} &= g_{\mu\nu} \left( {\cal M}_{ss} \right)^{1/6} + \gamma_{\alpha \beta} A_\mu{}^\alpha A_\nu{}^\beta \,, \\
  \hat{C}_{\mu\alpha\beta} &= A_\mu{}^s \epsilon_{\alpha\beta} \,, \\
  \hat{C}_{\mu\nu\alpha} &= \epsilon_{\alpha\beta} \left( - B_{\mu\nu}{}^{\beta,s} + A_{[\mu}{}^{\beta} A_{\nu]}{}^s \right) \\
  \hat{C}_{\mu\nu\rho} &= \epsilon_{\alpha\beta} \left( \frac12 C_{\mu\nu\rho}{}^{\alpha\beta,s} + A_{[\mu}{}^{\alpha} A_{\nu}{}^{\beta} A_{\rho]}{}^s - 3 A_{[\mu}{}^{\alpha} B_{\nu\rho]}{}^{\beta,s} \right) \,.
 \end{split} \label{eq:EFTM}
\end{equation}
Similarly, for IIB we have, with $\phi \equiv \phi_{ss}$ and the Kaluza-Klein vector $A_\mu{}^s = \phi^{-1} G_{\mu s}$
\begin{equation}
 \begin{split}
  {\cal M}_{ss} &= \phi^{8/7} \,, \\
  g_{\mu\nu} &= \phi^{1/7} \left( G_{\mu\nu} - \phi A_\mu{}^s A_\nu{}^s \right)   \,, \\
  A_\mu{}^\alpha &= \hat{C}_{\mu s}{}^\alpha \,, \\
  B_{\mu\nu}{}^{\alpha, s} &= \hat{C}_{\mu\nu}{}^{\alpha} + A_{[\mu}{}^s \hat{C}_{\nu]s}{}^\alpha \,, \\
  C_{\mu\nu\rho}{}^{\alpha\beta, s} &= \epsilon^{\alpha\beta} \hat{C}_{\mu\nu\rho s} + 3 \hat{C}_{[\mu|s|}{}^{[\alpha} \hat{C}_{\nu\rho]}{}^{\beta]} - 2 \hat{C}_{[\mu|s}{}^{\alpha} \hat{C}_{\nu|s|}{}^{\beta} A_{\rho]}{}^s \,, \\
  D_{\mu\nu\rho\sigma}{}^{\alpha\beta, ss} &= \epsilon^{\alpha\beta} \left( \hat{C}_{\mu\nu\rho\sigma} + 4 A_{[\mu}{}^s \hat{C}_{\nu\rho\sigma]s} \right) + 6 \hat{C}_{[\mu\nu}{}^{[\alpha} \hat{C}_{\rho|s|}{}^{\beta]} A_{\sigma]}{}^s \,,
 \end{split} \label{eq:IIBEFT}
\end{equation}
and 
\begin{equation}
 \begin{split}
  \phi &= \left( {\cal M}_{ss}\right)^{7/8} \,, \\
  G_{\mu\nu} &=  \left( {\cal M}_{ss} \right)^{-1/8} g_{\mu\nu} + \left(\gM_{ss} \right)^{7/8} A_\mu{}^s A_\nu{}^s \,, \\
  \hat{C}_{\mu s}{}^{\alpha} &= A_\mu{}^\alpha \,, \\
  \hat{C}_{\mu\nu}{}^{\alpha} &= B_{\mu\nu}{}^{\alpha,s} - A_{[\mu}{}^s A_{\nu]}{}^{\alpha} \,, \\
  \hat{C}_{\mu\nu\rho s} &= \frac12 \epsilon_{\alpha\beta} \left(C_{\mu\nu\rho}{}^{\alpha\beta,s} - 3 A_{[\mu}{}^{[\alpha} B_{\nu\rho]}{}^{\beta],s} - A_{[\mu}{}^{\alpha} A_{\nu}{}^{\beta} A_{\rho]}{}^s \right) \,, \\
  \hat{C}_{\mu\nu\rho\sigma} &= \frac12 \epsilon_{\alpha\beta} \left(D_{\mu\nu\rho\sigma}{}^{\alpha\beta,ss} + 6 B_{[\mu\nu}{}^{\alpha,s} A_{\rho}{}^{\beta} A_{\sigma]}{}^s - 4 A_{[\mu}{}^s C_{\nu\rho\sigma]}{}^{\alpha\beta,s} \right) \,.
 \end{split} \label{eq:EFTIIB}
\end{equation}

\section{Relationship to F-theory} 
\label{sec:Frel}

In the previous section, we have given the detailed rules for showing the equivalence of the $\G$ EFT to both 11-dimensional supergravity and 10-dimensional type IIB supergravity. Let us now elaborate on the connection to F-theory, rather than just type IIB supergravity.

What is F-theory? Primarily we will take F-theory to be a 12-dimensional lift of IIB supergravity that provides a geometric perspective on the $\mathrm{SL}(2)$ duality symmetry \footnote{We thank Cumrun Vafa for discussions on how one should think of F-theory.}. It provides a framework for describing (non-perturbative) IIB vacua with varying $\tau$, in particular it is natural to think of sevenbrane backgrounds as monodromies of $\tau$ under the action of $\mathrm{SL}(2)$. Equivalently, there is a process for deriving non-perturbative IIB vacua from M-theory compactifications to a dimension lower. Crucially, singularities of the 12-dimensional space are related to D7-branes. We take this duality with M-theory to be the second key property of F-theory.

Let us now briefly comment on how these two properties are realised here before expanding in detail.
We usually view the 12-dimensional space of F-theory as consisting of a torus fibration of 10-dimensional IIB. The group of large diffeomorphisms on the torus is then viewed as a geometric realisation of the $\mathrm{SL}(2)$ S-duality of IIB.

In the $\G$ EFT a similar picture arises. 
This is because we take the group of large \emph{generalised diffeomorphisms} acting on the extended space to give the $\G$ duality group. See \cite{Hohm:2012gk,Park:2013mpa, Berman:2014jba,Hull:2014mxa,Naseer:2015tia,Rey:2015mba,Chaemjumrus:2015vap} for progress on understanding the geometry of these large generalised diffeomorphisms.

The EFT is subject to a single constraint equation, the section condition, with two inequivalent solutions. One solution of the constraint leads to M-theory or at least 11-dimensional supergravity, and one leads to F-theory. Thus $\G$ EFT is a single 12-dimensional theory containing both 11-dimensional supergravity and F-theory, allowing us to naturally realise the M-theory / F-theory duality.

If we choose the IIB section, we can interpret any solutions as being 12-dimensional but with at least two isometries in the 12-dimensional space. These two isometries lead to the 2-dimensional fibration which in F-theory consists of a torus.

Finally, the fact that the generalised diffeomorphisms, not ordinary diffeomorphisms, play the key role here also allows one to use the section condition to ``dimensionally reduce'' the 12-dimensional $\G$ to 10-dimensional IIB (as well as 11-dimensional supergravity) as explained in section \ref{sec:supergravity}. This explicitly shows how F-theory, interpreted as the $\G$ EFT, can be a 12-dimensional theory, yet reduce to the correct 11-dimensional and IIB supergravity fields.

\subsection{M-theory/F-theory duality} \label{sec:MF}

Before looking at the specifics of the $\G$ EFT, let us discuss in general how duality works in exceptional field theories or indeed double field theory. First consider the most familiar case of T-duality and how it comes about in DFT. The origin of T-duality is the ambiguity in identifying the $d$-dimensional spacetime embedded in the $2d$-dimensional doubled space. For the most generic DFT background that obeys the section condition constraint there is no T-duality. The section condition eliminates the dependence of the generalised metric on half the coordinates and so there is unique choice of how one identifies the $d$-dimensional spacetime embedding in the $2d$-doubled space.

However if there is an isometry then there is only a dependence on $d-1$ coordinates. Thus what we identify as the $d$-dimensional spacetime is ambiguous. This ambiguity is T-duality. (Note, even though DFT has a manifest $O(d,d;\mathbb{R})$ local symmetry this should not be confused with the global $O(n,n;\mathbb{Z})$ T-duality which only occurs when compactifying on a $T^n$.)

One has a similar situation in EFT but with some differences. In DFT a solution to the section condition will always provide a $d$-dimensional space. In EFT a solution to the $\Edd$ section condition will provide either a $d$-dimensional space or a $\left(d-1\right)$-dimensional space (where crucially the $d-1$ solution is not a subspace of the $d$-dimensional space). The two solutions are distinct (and not related by any element of $\Edd$). The $d$-dimensional solution is associated to the M-theory description and the $\left(d-1\right)$-dimensional solution is associated to the type IIB description. 

A completely generic solution that solves the section condition will be in one set or the other and one will be able to label it as an M or IIB solution. However, if there are {\bf{two}} isometries in the M-theory solution then again we have an ambiguity and one will be able interpret the solution in terms of IIB section with {\bf{one}} isometry. This ambiguity gives the F-theory/M-theory duality. It is the origin of how M-theory on a torus is equivalent to IIB on a circle \cite{Schwarz:1995jq}. Thus in summary the F/M-duality is an ambiguity in the identification of spacetime that occurs when there are two isometries in an M-theory solution.

So the simplest case is where the eleven dimensional space is a $M_9 \times T^2$ and we take the 2-torus to have complex structure $\tau$ and volume $V$. The  reinterpretation in terms of a IIB section solution (which requires a single circle isometry) is given by:
\be
R^{IIB}_9 = V^{-3/4}
\eeq
where $R^{IIB}_9$ is the radius of the IIB circle. This exactly recovers the well known M/IIB duality relations \cite{Schwarz:1995jq}.

Let us now further study the M-theory / F-theory duality with the following simple Ansatz. Consider the 9-dimensional space to be of the form
\begin{equation}
 \dd s_{(9)}^2 = \dd s_{(1,2)}^2 + \dd s_{(6)}^2 \,,
\end{equation}
where $\dd s_{(6)}^2$ is the metric on some internal 6-dimensional manifold $B_6$ and $\dd s_{(1,2)}^2$ is the metric of an effective three-dimensional theory.
We will consider the case where $y^\alpha$ parameterise a torus, as in F-theory. If for simplicity we ignore the one-form gauge potential $\Aa_\mu{}^M$ then using \eqref{eq:EFTM} we see that the M-theory section has the following metric
\begin{equation}
 \dd s_{(11)}^2 = \left({\cal M}_{ss}\right)^{1/6} \dd s_{(1,2)}^2 
 		+ \left[ \left( {\cal M}_{ss} \right)^{1/6} \dd s_{(6)}^2 
 		+ \left({\cal M}_{ss}\right)^{-7/12} {\cal H}_{\alpha\beta} \dd y^\alpha \dd y^\beta \right] \,.
\end{equation}
The internal manifold takes the form of a $T^2$-fibration over $B_6$. In the IIB section we instead have the Einstein-frame metric
\begin{equation}
 \dd s_{(10)}^2 = \left[ \left( {\cal M}_{ss} \right)^{-1/8} \dd s_{(1,2)}^2 
 	+ \left( {\cal M}_{ss} \right)^{7/8} (\dd y^s)^2 \right]
 	+ \left( {\cal M}_{ss} \right)^{-1/8} \dd s_{(6)}^2 \,.
\end{equation}
This is precisely the four-dimensional effective theory with six-dimensional internal space that we expect from F-theory, with the fourth direction $y^s$ becoming ``large'' in the small-volume limit, here given by ${\cal M}_{ss} \rightarrow \infty$. The dilaton and $C_0$ profile are given by ${\cal H}_{\alpha\beta}$ and are at this point arbitrary.

Let us mention another example of this M-theory / IIB relationship which will be important to us later. Consider M-theory with a three-form $\hat{C}_{tx^1x^2}$. It is easy to show from the dictionary \eqref{eq:MEFT} -- \eqref{eq:EFTIIB} that in IIB this leads to a 4-form tangential to the 4-dimensional spacetime $\hat{C}_{tx_1x_2s}$.

\subsection{Sevenbranes}

In F-theory, a vital role is played by backgrounds containing sevenbranes. In this subsection we discuss some features of how one may view sevenbranes and their singularities in the context of the $\G$ EFT.

Sevenbrane solutions of type IIB supergravity have non-trivial metric and scalar fields $\tau$. From the point of view of EFT, all of these degrees of freedom are contained within the metric $g_{\mu \nu}$ and the generalised metric $\gM_{MN}$. Thus we may specify entirely a sevenbrane background by giving these objects. In the below, we will use the notation
\begin{align}
\dd s^2_{(9)} &= g_{\mu\nu}\dd x^\mu \dd x^\nu \, , &
\dd s^2_{(3)} &= \MM_{\alpha\beta}\dd y^\alpha \dd y^\beta + \MM_{ss} ( \dd y^s)^2\, ,
\end{align}
to specify the solutions. It is not obvious that one should view the generalised metric as providing a notion of line element on the extended space, so in a sense this is primarily a convenient shorthand for expression the solutions.

We consider a sevenbrane which is extended along six of the ``external directions'', denoted $\vec{x}_6$, and along $y^s$ which appears in the extended space. The remaining coordinates are time and the coordinates transverse to the brane which we take to be $(r,\theta)$ polar coordinates. In this language, the harmonic function of the brane is $H \approx h \ln[r_0/r]$.\footnote{In the EFT point of view, one can consider the sevenbrane as having the form of a generalised monopole solution of the form given in section \ref{sec:monopole}, after smearing on one of the transverse directions. In this case $r_0$ is a cut-off that is introduced in this process and is related to the codimension-2 nature of the solution, i.e. we expect it to be valid only up to some $r_0$ as the solution is not asymptotically flat.} The solution can be specified by
\begin{equation}
\begin{aligned}
\dd s^2_{(9)} &= -\dd t^2 + \dd \vec{x}_{(6)}^{\, 2} 
						+ H \left(\dd r^2 + r^2 \dd \theta^2 \right)  \\
\dd s^2_{(3)} &= H^{-1}\left[(\dd y^1)^2 + 2 h\theta \dd y^1\dd y^2 + K (\dd y^2)^2 \right]
						+ (\dd y^s)^2  \\
{\Aa_\mu}^M &= 0 \, , \qquad K = H^2 + h^2 \theta^2 \, .
\end{aligned}
\label{eq:sevenbrane}
\end{equation}
If one goes around this solution in the transverse space changing $\theta=0$ to $\theta=2\pi$ the $2\times 2$ block $\MM_{\alpha\beta}$ goes to
\begin{equation}
\MM \rightarrow \Omega^T\MM\Omega
\end{equation}
where the monodromy $\Omega$ is an element of $\mathrm{SL}(2)$ and is given by 
\begin{equation}
\Omega = 
\begin{pmatrix}
1 & 2\pi h \\ 0 & 1 
\end{pmatrix} \, .
\end{equation}

Reducing this solution to the IIB section gives the D7-brane. By using \eqref{eq:EFTIIB} one can extract the torus metric $\cH_{\alpha\beta}$ and the scalar $\phi$ of the $10=9+1$ split. From $\cH_{\alpha\beta}$ one then obtains the axio-dilaton, i.e. $C_0$ and $e^\varphi$. The external metric is composed with $\phi$ to give the 10-dimensional solution
\begin{equation}
\begin{aligned}
\dd s^2_{(10)} &= -\dd t^2 + \dd \vec{x}_{(6)}^{\, 2} 
					+ H \left(\dd r^2 + r^2 \dd \theta^2 \right)	+ (\dd y^s)^2  \\
C_0 &= h\theta \, , \qquad e^{2\varphi} = H^{-2}			
\end{aligned}
\label{eq:D7}
\end{equation}
which is the D7-brane. Exchanging $y^1$ and $y^2$ and flipping the sign of the off-diagonal term (this is an $\mathrm{SL}(2)$ transformation of the $\MM_{\alpha\beta}$ block) leads to a solution which reduces to the S7-brane. On the M-theory section the solution \eqref{eq:sevenbrane} corresponds to a smeared KK-monopole which can be written as
\begin{equation}
\dd s^2_{(11)} = -\dd t^2 + \dd \vec{x}_{(6)}^{\, 2} 
					+ H \left(\dd r^2 + r^2 \dd \theta^2	+ (\dd y^1)^2\right)
					+ H^{-1}\left[\dd y^2 + h\theta\dd y^1\right]^2\, .
\label{eq:smearedKK}
\end{equation}
To see this more clearly, consider the usual KK-monopole in M-theory, which has three transverse and one isometric direction, the Hopf fibre. If this solution is smeared over one of the transverse directions to give another isometric direction one arrives at the above solution (where $(r,\theta)$ are transverse and $(y^1,y^2)$ are isometric). Therefore the M/IIB-duality between smeared monopole and sevenbrane relates the first Chern class of the Hopf fibration to the monodromy of the codimension-2 object. 

We have now seen how the $\mathrm{SL}(2)$ doublet of D7 and S7 is a smeared monopole with its two isometric direction along the $y^\alpha$ in the extended space. This can be generalized to give $pq$-sevenbranes in the IIB picture where the isometric directions of the smeared monopole correspond to the $p$- and $q$-cycles. The external metric is the same as above, the generalized metric now reads
\begin{equation}
\begin{aligned}
\dd s^2_{(3)} &= \frac{H^{-1}}{p^2+q^2}\bigg\{
					\left[p^2H^2 + (ph\theta - q)^2\right](\dd y^1)^2 
					+ \left[(p + qh\theta)^2 + q^2H^2\right](\dd y^2)^2  \\
	&\hspace{3cm} - 2\left[(p^2-q^2)h\theta + pq(K-1)\right] \dd y^1\dd y^2 \bigg\}
					+  (\dd y^s)^2  \, .
\end{aligned}
\label{eq:pqsevenbrane}
\end{equation}
The two extrema are $p=0$ which gives the D7 and $q=0$ which gives the S7. 

As for all codimension-2 objects, a single D7-brane should not be considered on its own. To get a finite energy density, a configuration of multiple sevenbranes needs to be considered. Introducing the complex coordinate $z$ on the two-dimensional transverse space, such a multi-sevenbrane solution in EFT reads
\begin{equation}
\begin{aligned}
\dd s^2_{(9)} &= -\dd t^2 + \dd \vec{x}_{(6)}^{\, 2} + \tau_2|f|^2\dd z \dd \bz  \\
\dd s^2_{(3)} &= \frac{1}{\tau_2}\left[|\tau|^2 (\dd y^1)^2 
						+ 2\tau_1 \dd y^1\dd y^2 + (\dd y^2)^2 \right]	+(\dd y^s)^2 
\end{aligned}
\label{eq:multisevenbrane}
\end{equation}
where all the tensor fields still vanish. Instead of specifying a harmonic function on the transverse space, we now have the holomorphic functions $\tau(z)$ and $f(z)$. Their poles on the $z$-plane correspond to the location of the sevenbranes \cite{Greene:1989ya, Gibbons:1995vg}. One usually takes 
\be
\tau = j^{-1} \left( \frac{P(z)}{Q(z)} \right) \,,
\ee
where $P(z)$ and $Q(z)$ are polynomials in $z$. The roots of $Q(z)$ will give singularities, which in the IIB section give the locations of the sevenbranes. The configuration in this case consists of the metric 
\be
\dd s^2_{(10)} = -\dd t^2 + \dd \vec{x}_{(6)}^{\, 2} + \dd y_s^2 + \tau_2|f|^2\dd z \dd \bz  
\ee
together with the scalar fields encoded by $\tau$. Meanwhile in the M-theory section, one finds a purely metric background,
\be
\dd s^2_{(11)}  = - \dd t^2 + \dd \vec{x}_{(6)}^{\, 2}
+ \tau_2|f|^2\dd z \dd \bz 
 + \tau_2 ( \dd y^1 )^2 
+ \frac{1}{\tau_2} \left( \dd y^2 + \tau_1 \dd y^1 \right)^2 \,.
\ee
This retains the singularities at the roots of $Q$, at which $\tau_2 \rightarrow i \infty$. 

A crucial point about singularities in F-theory is that they are seen as an origin for nonabelian gauge symmetries. These arise from branes wrapping vanishing cycles at the singularities. At this point we will not examine the details of this for our $\G$ EFT but instead we can point to interesting recent work in DFT. The authors of \cite{Aldazabal:2015yna} construct the full nonabelian gauge enhanced theory in DFT corresponding to the bosonic string at the self-dual point. One would hope that one could apply a simlar construction to produce a gauge enhanced EFT with non-abelian massless degrees of freedom coming from wrapped states on vanishing cycles.

\section{Solutions}
\label{sec:solutions} 

In this section, we discuss the embedding of supergravity solutions into the $\G$ EFT, in particular showing how one can use the EFT form of a configuration to relate 11- and 10-dimensional solutions. 

\subsection{Membranes, strings and waves}
\label{sec:waves}

In higher rank EFT constructions, and double field theory, the fundamental string and the M2 solutions appear as generalised wave solutions in the extended space \cite{Berkeley:2014nza}. This applies in the case when all worldvolume spatial directions of these branes lie in the extended space. In this subsection, we will describe first the form of such solutions in the $\G$ EFT, and then also look at the case where the extended space is transverse to an M2.

\subsubsection{Waves in extended space} 

The first family of solutions we are considering can be thought of as a null wave in EFT.\footnote{To properly think of a given solution as being a wave carrying momentum in a particular direction of the extended space, one should construct the conserved charges associated to generalised diffeomorphism invariance, as has been done in the DFT case \cite{Blair:2015eba , Park:2015bza , Naseer:2015fba}. } This can be reinterpreted on a solution of the section condition as branes whose worldvolume spatial directions are wholly contained in the extended space. We denote the spatial coordinates of the external space by $\vec{x}_8$, which become the transverse directions of the supergravity solutions. The harmonic function that appears is $H=1+\frac{h}{|\vec{x}_{(8)}|^6}$, $h$ a constant. As the fields only depend on $\vec{x}_{(8)}$ the section condition \eqref{eq:constraint} is automatically satisfied. The external metric and the one-form field for this set of solutions are
\begin{equation}
\begin{aligned}
\dd s^2_{(9)} &= H^{1/7}\left[-H^{-1}\dd t^2 + \dd \vec{x}_{(8)}^{\, 2}\right]  \\
{\Aa_t}^M &= -(H^{-1}-1)a^M \, .
\end{aligned}
\label{eq:waveexternal}
\end{equation}
Here $a^M$ is a unit vector in the extended space which -- depending on its orientation -- distinguishes between the different solutions in this family. For $a^M=\delta^M_s$, i.e. a wave propagating along $y^s$, the generalized metric is
\begin{equation}
\dd s^2_{(3)} =   H^{-6/7}\left[\delta_{\alpha\beta}\dd y^\alpha \dd y^\beta 
						+ H^2(\dd y^s)^2 \right]  \, .
\label{eq:membrane}
\end{equation}
Upon a reduction to the M-theory section this corresponds to a membrane (M2) stretched along $y^1$ and $y^2$. On the IIB section one obtains a pp-wave propagating along $y^s$ . In other words, a membrane wrapping the two internal directions is a wave from a IIB point of view.

Explicitly, one sees that the internal two-metric $\gamma_{\alpha\beta}$ of the $11=9+2$ split of M-theory together with its determinant can be extracted from the above line element with the help of \eqref{eq:genmetricM} as
\begin{align}
\gamma_{\alpha\beta} &= H^{-2/3}\delta_{\alpha\beta} \, , & \gamma &= H^{-4/3} \, .
\end{align}
Using the dictionary \eqref{eq:EFTM} we obtain the membrane solution
\begin{equation}
\begin{aligned}
\dd s^2_{(11)} &= H^{-2/3}\left[-\dd t^2 + \delta_{\alpha\beta}\dd y^\alpha \dd y^\beta \right]
					+ H^{1/3} \dd \vec{x}_{(8)}^{\, 2} \\
C_{ty^1y^2} &= -(H^{-1}-1) \, .			
\end{aligned}
\label{eq:M2}
\end{equation}
A similar procedure on the IIB section using \eqref{eq:genmetricIIB} and \eqref{eq:EFTIIB} leads to the pp-wave. This time the EFT vector ${\Aa_t}^s$ yields the KK-vector in the $10=9+1$ split, and one gets the pp-wave metric
\begin{equation}
\begin{aligned}
\dd s^2_{(10)} &= -H^{-1}\dd t^2 + H\left[\dd y^s -(H^{-1}-1) \dd t\right]^2
					+  \dd \vec{x}_{(8)}^{\, 2} \, .			
\end{aligned}
\label{eq:pp}
\end{equation}
If on the other hand the solution \eqref{eq:membrane} is oriented along one of the $y^\alpha$, say $y^1$ and thus $a^M=\delta^{M}_{1}$, the corresponding generalized metric is
\begin{equation}
\dd s^2_{(3)} = H^{9/14} \left[H^{1/2}(\dd y^1)^2 + H^{-1/2}(\dd y^2)^2 
					+ H^{-3/2} (\dd y^s)^2 \right] \, .
\label{eq:string}					
\end{equation}
This now corresponds to a fundamental string on the IIB section. The opposite orientation, i.e. $a^M=\delta^{M}_{ 2}$ and swapping $y^1$ and $y^2$ in $\MM_{MN}$, gives the D1-brane. The reduction is performed as explained above, resulting in the 10-dimensional solutions
\begin{equation}
\begin{aligned}
\dd s^2_{(10)} &= H^{-3/4}\left[-\dd t^2 + (\dd y^s)^2 \right]
					+ H^{1/4} \dd \vec{x}_{(8)}^{\, 2} \\
\mathrm{F1:} \quad B_{ts} &= -(H^{-1}-1) \\
\mathrm{D1:} \quad C_{ts} &= -(H^{-1}-1) \\
e^{2\varphi} &= H^{\mp 1} \, .			
\end{aligned}
\label{eq:FD1}
\end{equation}
Here $e^{2\varphi}$ is the string theory dilaton. The fundamental string comes with the minus sign in the dilaton and couples to the NSNS-two-form $B_{ty^s}$. The D-string has the plus sign in the dilaton and couples to the RR-two-form $C_{ty^s}$.

On the M-theory section the solution \eqref{eq:string} corresponds to a pp-wave propagating along $y^1$ or $y^2$. Thus the $\mathrm{SL}(2)$ doublet of F1 and D1 is a wave along the $y^\alpha$ directions in the extended space. This can be generalized to give the $pq$-string in the IIB picture by orienting the wave in a superposition of the two $y^\alpha$ directions like $\frac{py^1+qy^2}{\sqrt{p^2+q^2}}$. Then the direction vector is given by
\begin{equation}
a^M = \frac{1}{\sqrt{p^2+q^2}}
\begin{pmatrix}
p \\ q \\ 0
\end{pmatrix}
\label{eq:pqvector}
\end{equation}
and the generalized metric for this configuration is
\begin{equation}
\begin{aligned}
\dd s^2_{(3)} &= H^{9/14}\left\{\frac{H^{-1/2}}{p^2+q^2}\left[(p^2H+q^2)(\dd y^1)^2 
					+ 2pq(H-1)\dd y^1 \dd y^2 \right.\right. \\
				&\hspace{4.5cm}\left.\left.
					+ (p^2+q^2H)(\dd y^2)^2 \right] + H^{-3/2} (\dd y^s)^2 
					\vphantom{\frac{H^{-1/2}}{p^2+q^2}}\right\}
\end{aligned}
\label{eq:pqstring}
\end{equation}
which reduces to the F1 for $q=0$ and the D1 for $p=0$. On the M-theory section this now corresponds to a wave propagating in a superposition of the two internal directions. 

The solutions \eqref{eq:membrane}, \eqref{eq:string} and \eqref{eq:pqstring} form a family which all look the same from the external point of view and only differ in the generalised metric (and the corresponding orientation in the extended space via $a^M$). The EFT wave solutions therefore nicely combines the membrane solution of M-theory and the F1, D1 and $pq$-string solutions of IIB into a single mode propagating in the extended space.

\subsubsection{M2 and D3} 

We have now seen an M2-brane wrapping the $y^\alpha$ space maps to a pp-wave in the IIB theory. In this case, as we mentioned, the worldvolume directions were aligned within the extended space and in particular this meant that we automatically had isometries in these directions. 

Now instead consider an M2-brane extending on two of the external directions plus time, so that its worldvolume extends along $t$ and $\vec{x}_{(2)}$. From the general considerations of the previous section, we expect this to get mapped to a D3-brane along the $t$, $\vec{x}_{(2)}$ and $y^s$ directions. Let's see this explicitly. In these coordinates, the supergravity solution of the M2-brane is given by
\begin{equation}
\begin{aligned}
\dd s^2_{(11)} &= H^{-2/3}\left[-\dd t^2 + \dd \vec{x}_{(2)}^{\, 2} \right]
					+ H^{1/3} \left[\dd \vec{w}_{(6)}^{\, 2} 
						+ \delta_{\alpha\beta}\dd y^\alpha \dd y^\beta \right] \\
\hat C_{tx^1x^2} &= -(H^{-1}-1) \, .			
\end{aligned}
\end{equation}
The harmonic function is a function of the transverse coordinates $H = H(\vec{w}_{(6)}, y^\alpha)$. We take the $y^\alpha$ to be compact so that the harmonic function must be periodic in $y^1$ and $y^2$. With respect to the 9-dimensional theory, the dependence on the $y^\alpha$ coordinates gives rise to massive fields with masses inversely proportional to the volume of the $y^\alpha$ space. In the usual ``F-theory limit'' where this volume is taken to zero, we can thus ignore the dependence on $y^1$ and $y^2$ and smear the solution over the $y^\alpha$ space. Using the dictionary \eqref{eq:MEFT}, we see that as an EFT solution this would correspond to
\begin{equation}
\begin{aligned}
\dd s^2_{(9)} &= H^{-4/7}\left[-\dd t^2 + \dd \vec{x}_{(2)}^{\, 2}\right] 
					+ H^{3/7} \dd \vec{w}_{(6)}^{\, 2}   \\
\dd s^2_{(3)} &= H^{3/7}\delta_{\alpha\beta}\dd y^\alpha \dd y^\beta	+ H^{-4/7} (\dd y^s)^2 \\
{\Ac_{tx^1x^2}}^{[\alpha\beta]s} &= \epsilon^{\alpha\beta}\hat C_{tx^1x^2} \, .
\end{aligned}
\end{equation}

We find that on the IIB section this corresponds to a D3-brane wrapped on the 4-dimensional spacetime $\dd s_{(4)}^2 = -\dd t^2 + \dd \vec{x}_{(2)}^{\, 2} + (\dd y^s)^2$, as one would expect. In particular, we have
\begin{equation}
\begin{aligned}
\dd s^2_{(10)} &= H^{-1/2} \left[-\dd t^2 + \dd \vec{x}_{(2)}^{\, 2} + (\dd y^s)^2 \right] 
					+ H^{1/2} \dd \vec{w}_{(6)}^{\, 2}  \\
\hat{C}_{tx^1x^2s} &= -(H^{-1}-1) \,, \qquad e^{\varphi} = 1 \,.
\end{aligned}
\end{equation}

\subsection{Fivebranes and monopoles}
\label{sec:monopole}

Similar to the relation between generalized null waves, strings and membranes above, the solitonic fivebrane and the M5 solutions can be seen to appear as generalised monopole solutions in the extended space \cite{Berman:2014jsa}, in the case where the special isometry direction of the monopole lies in the extended space. We will first review this case, and then give the other solutions in which one obtains M5 branes (partially) wrapping the internal space. 

\subsubsection{Monopole in extended space} 

The first set of solutions we consider takes the form of a monopole structure (Hopf fibration) in the extended space, so as stated above the special isometric direction of the monopole is one of the $Y^M$. The coordinates on the external space are denoted $x^\mu = ( t, \vec{x}_{(5)}, \vec{w}_{(3)})$. The harmonic function is $H=1+\frac{h}{|\vec{w}_{(3)}|}$ (satisfying the section condition) and the field configuration of the common sector is
\begin{equation}
\begin{aligned}
\dd s^2_{(9)} &= H^{-1/7}\left[-\dd t^2 + \dd \vec{x}_{(5)}^{\, 2} 
						+ H \dd \vec{w}_{(3)}^{\, 2}\right]  \\
{\Aa_i}^M &= A_ia^M \, , \qquad 2\partial_{[i}A_{j]} = {\epsilon_{ij}}^k\partial_kH \, .
\end{aligned}
\label{eq:monopoleexternal}
\end{equation}
Here $A_i$ is the magnetic potential obtained by dualising the 6-form ${\Af_{tx^1\dots x^5}}^M = -(H^{-1}-1)a^M$ (and $w^i$ with $i=1,2,3$ are the transverse coordinates). As for the EFT wave, the unit vector $a^M$ distinguishes between the solutions in this family by specifying the isometric direction. If the monopole is oriented along $y^s$ and therefore $a^M=\delta^M_s$, the generalized metric is
\begin{equation}
\dd s^2_{(3)} = H^{6/7}\left[\delta_{\alpha\beta}\dd y^\alpha \dd y^\beta 
						+ H^{-2} (\dd y^s)^2 \right] \, .
\label{eq:fivebraneM}
\end{equation}
This corresponds to a fivebrane (M5) on the M-theory section. Note that the fivebrane is smeared over two of its transverse directions. Using the dictionary \eqref{eq:EFTM}, the internal two-metric extracted from this generalized metric combines with the external metric to give an 11-dimensional solution with the $C_{iy^1y^2}$ component of the M-theory three-form provided by ${\Aa_i}^s$. This can be dualised to give the six-form the fivebrane couples to electrically. The full solution reads
\begin{equation}
\begin{aligned}
\dd s^2_{(11)} &= H^{-1/3}\left[-\dd t^2 + \dd \vec{x}_{(5)}^{\, 2} \right]
					+ H^{2/3} \dd \vec{w}_{(3)}^{\, 2} \\
C_{iy^1y^2} &= A_i \, , \qquad C_{tx^1\dots x^5} = -(H^{-1}-1) \, .			
\end{aligned}
\label{eq:M5}
\end{equation}
If on the other hand \eqref{eq:fivebraneM} is considered on the IIB section, one obtains the KK-monopole in ten dimensions via \eqref{eq:EFTIIB} where the KK-vector for the $10=9+1$ split is encoded in  ${\Aa_i}^s$. The 10-dimensional solution is given by
\begin{equation}
\dd s^2_{(10)} = -\dd t^2 + \dd \vec{x}_{(5)}^{\, 2} + H^{-1}\left[\dd y^s + A_i\dd w^i\right]^2
					+ H \dd \vec{w}_{(3)}^{\, 2} \, .			
\label{eq:KK6B}
\end{equation}

The alternative choice for the isometric direction of the EFT monopole is one of the $y^\alpha$, say $y^1$ and thus $a^M=\delta^{M}_{ 1}$, the corresponding generalized metric is
\begin{equation}
\dd s^2_{(3)} = H^{-9/14}\left[H^{-1/2}(\dd y^1)^2 + H^{1/2}(\dd y^2)^2 
					+ H^{3/2} (\dd y^s)^2 \right] \, .
\label{eq:fivebraneIIB}
\end{equation}
This is the solitonic fivebrane (NS5) in the IIB picture. Note again that the resulting fivebrane is smeared over one of its transverse directions which corresponds to the isometric direction of the monopole. One can also pick the opposite choice for $a^M$, i.e. $a^M=\delta^{M}_{ 2}$, which gives the D5-brane. The reduction procedure should be clear by now, the 10-dimensional solution is
\begin{equation}
\begin{aligned}
\dd s^2_{(10)} &= H^{-1/4}\left[-\dd t^2 + \dd \vec{x}_{(5)}^{\, 2} \right]
					+ H^{3/4} \left[\dd \vec{w}_{(3)}^{\, 2} + (\dd y^s)^2 \right] \\
\mathrm{NS5:} \quad B_{is} &= A_i   	\qquad 	B_{tx^1\dots x^5} = -(H^{-1}-1) \\
\mathrm{D5:} \quad C_{is} &= A_i 		\qquad  C_{tx^1\dots x^5} = -(H^{-1}-1) \\
e^{2\varphi} &= H^{\pm 1}			
\end{aligned}
\label{eq:NSD5}
\end{equation}
where the NS5 couples to the NSNS-B-fields and takes the positive sign in the dilaton while the D5 couples to the RR-C-fields and takes the negative sign in the dilaton. So the $\mathrm{SL}(2)$ doublet of NS5 and D5 is a monopole with its isometric direction along the $y^\alpha$ in the extended space. On the M-theory section the solution \eqref{eq:fivebraneIIB} is a KK-monopole with its isometric direction along one of the two internal dimensions.

As for the EFT wave, this can be generalized to give the $pq$-fivebrane of the IIB picture by having the isometric direction in a superposition of the two $y^\alpha$ directions like $\frac{py^1+qy^2}{\sqrt{p^2+q^2}}$. Then the $a^M$ is given again by \eqref{eq:pqvector} and the generalized metric for this configuration is
\begin{equation}
\begin{aligned}
\dd s^2_{(3)} &= H^{-9/14}\left\{\frac{H^{1/2}}{p^2+q^2}\left[(p^2H^{-1}+q^2)(\dd y^1)^2 
					+ 2pq(H^{-1}-1) \dd y^1\dd y^2 \right.\right. \\
				&\hspace{5.5cm}\left.\left.
					+ (p^2+q^2H^{-1})(\dd y^2)^2 \right] + H^{3/2} (\dd y^s)^2 
					\vphantom{\frac{H^{1/2}}{p^2+q^2}}\right\} \, .
\end{aligned}
\label{eq:pqfivebrane}
\end{equation}
For $q=0$ this is the NS5 and for $p=0$ this gives the D5 (both smeared). On the M-theory section this now corresponds to a KK-monopole with its isometric direction in a superposition of the two internal directions.

The EFT monopole combines the fivebrane solution of M-theory and the NS5, D5 and $pq$-fivebrane solutions of IIB into a single monopole structure with isometric direction in the extended space.

\subsubsection{M5 wrapped on internal space}

In the above we saw that M5-branes oriented completely along the external nine directions lead to KK-monopoles in the IIB picture. Let us now also study M5-branes (partially) wrapping the $y^\alpha$ space. We consider the setup in section \ref{sec:MF} where our external spacetime consists of a 3-dimensional part containing time and with spatial coordinates $\vec{x}_{(2)}$ and a 6-dimensional part denoted $B_6$.

Let us begin with an M5-brane wrapping both $y^1$ and $y^2$, as well as wrapping a 2-cycle $\Sigma_2$ in $B_6$ (along the two directions $\vec{z}_{(2)}$) and being extended along $t$ and $x^1$. This gives a D3-brane extended along $t$, $x^1$ and wrapping the 2-cycle $\Sigma_2$, as expected. The 12-dimensional EFT solution to which this corresponds is given by
\begin{equation}
 \begin{split}  
 \dd s_{(9)}^2 &= H^{-3/7} \left[-\dd t^2 + (\dd x^1)^2 + \dd \vec{z}_{(2)}^{\, 2} \right] 
  						+ H^{4/7} \left[(\dd x^2)^2 + \dd \vec{w}_{(4)}^{\, 2}\right] \,, \\
  \dd s_{(3)}^2 &= H^{-2/21} \delta_{\alpha\beta}\dd y^\alpha \dd y^\beta 
  						+ H^{4/7} (\dd y^s)^2 \,,
 \end{split}
\end{equation}
with non-trivial gauge field $\Ac_{\mu\nu\rho}{}^{[\alpha\beta]s}$. On the M-theory section this gives rise to the solution
\begin{equation}
\begin{aligned}
\dd s^2_{(11)} &= H^{-1/3}\left[-\dd t^2 + (\dd x^1)^2 + \delta_{\alpha\beta}\dd y^\alpha \dd y^\beta
			+	\dd \vec{z}_{(2)}^{\, 2}	  \right]
			+ H^{2/3} \left[(\dd x^2)^2 + \dd \vec{w}_{(4)}^{\, 2}\right] \, ,	
\end{aligned}
\end{equation}
with 3-form $\hat{C}_3$ which couples magnetically to the M5. 
In the IIB section we find the expected D4-brane
\begin{equation}
\dd s^2_{(10)} = H^{-1/2} \left[-\dd t^2 + (\dd x^1)^2 +	\dd \vec{z}_{(2)}^{\, 2}	 \right] 
			+ H^{1/2} \left[ (dy^s)^2 + (\dd x^2)^2 + \dd \vec{w}_{(4)}^{\, 2} \right] \,,
\end{equation}
with magnetic 4-form $\hat{C}_{\mu\nu\rho s} = \left(\hat{C}_{3}\right)_{\mu\nu\rho}$ and all other fields vanishing.

We can also consider an M5-brane which wraps a closed 3-cycle $\Sigma_3$ of $B_6$ as well as one of the $y^{\alpha}$. In the IIB picture this should correspond to a NS5-brane or D5-brane, for $\alpha = 1$ or $2$. Let us denote the coordinates on $B_6$ as $\vec{z}_{(3)}$ for the $\Sigma_3$ and $\vec{w}_{(3)}$ for the other three coordinates. The explicit EFT solution is given by
\begin{equation}
 \begin{split}  
 ds_{(9)}^2 &= H^{-2/7} \left[-\dd t^2 + (\dd x^1)^2 + \dd \vec{z}_{(3)}^{\, 2} \right] 
 			+ H^{5/7} \left[ (\dd x^2)^2 + \dd \vec{w}_{(3)}^{\, 2} \right] \, , \\
  ds_{(3)}^2 &= H^{-2/7} (\dd y^1)^2 + H^{5/7} (\dd y^2)^2 + H^{-2/7} (dy^s)^2 \,, 
 \end{split}
\end{equation}
and gauge field $B_{\mu\nu}{}^{\alpha s} = \epsilon^{\alpha\beta} \hat{C}_{\mu\nu\beta}$ where $\hat{C}_{\mu\nu\alpha}$ will be specified shortly. As for the M2/D3 case, we can in the zero-volume limit take the harmonic functions to be smeared across the $y^1$ direction. On the M-theory section this gives the expected M5-brane
\begin{equation}
\dd s^2_{(11)} = H^{-1/3} \left[-\dd t^2 + (\dd x^1)^2 + (\dd y^1)^2 
					+ \dd \vec{z}_{(3)}^{\, 2} \right] 
					+ H^{2/3} \left[(\dd x^2)^2 + \dd \vec{w}_{(3)}^{\, 2} + (dy^s)^2 \right] \,.
\end{equation}
The gauge field has only non-zero components $\hat{C}_{z^iz^jy^2}$, where $i = 1, \ldots, 3$. On the other hand, on the IIB section we obtain a D5-brane along $t$, $x^1$ and wrapping $y^s$ as well as $\Sigma_3$
\begin{equation}
\dd s^2_{(10)} = H^{-1/4} \left[-\dd t^2 + (\dd x^1)^2  + (dy^s)^2 
					+ \dd \vec{z}_{(3)}^{\, 2}  \right]
					+ H^{3/4} \left[ (\dd x^2)^2 + \dd \vec{w}_{(3)}^{\, 2} \right] \,.	
\end{equation}
The 6-form gauge field is then the dual to $\hat{C}_{z^{i}z^{j}} = \hat{C}_{z^{i} z^{j} y^2}$ and we find a D5-brane. By an $\mathrm{SL}(2)$ transformations we can also obtain NS5-branes this way.

\subsection{Comment on higher rank duality} 

We saw above that the solutions described in section \ref{sec:waves} in which one had waves in the extended space (corresponding to F1, D1 and M2 totally wrapping the internal space) essentially split into two categories, depending on whether the vector $a^M$ giving the wave's direction pointed in the $y^\alpha$ or $y^s$ directions. These different solutions were not related by duality.\footnote{ However they can be related by a $Z_2$ transformation on the generalised metric and the extended coordinates as in \cite{Malek:2015hma}. This transformation maps the common NS-NS sector of type IIA / IIB into each other and would thus allow us to map the generalised metric containing the membrane into the F1-string in IIB.} However, in higher rank duality groups one finds that all branes whose worldvolumes are only spatially extended in the internal space are unified into a single solution of the EFT \cite{Berman:2014hna,Berman:2014jsa}. Similarly, as one increases the rank of the duality group considered, various of the M2 and M5 brane solutions will be unified as more of the worldvolume directions fall into the internal space. 
Thus, families of these solutions appear in a unified manner. With this lesson learned, one may hope that studying F-theory compactifications in bigger EFTs may allow one to more easily consider complicated set-ups.

\section{Conclusions and outlook}

We are aware that the approach of most practitioners in F-theory that has yielded so much success over a number of years has been through algebraic geometry. It is doubtful if the presence of this action can help in those areas where the algebraic geometry has been so powerful. We do hope though that it may provide some complementary techniques given that we now have a description in terms of 12-dimensional degrees of freedom equipped with an action to determine their dynamics. 

One question people have tried to answer is the theory on a D3-brane when $\tau$ varies. This might be computable in this formalism using a Goldstone mode type analysis similar to that in  \cite{Berkeley:2014nza} where a Goldstone mode analysis was used to determine string and brane effective actions in DFT and EFT. 

A useful result from this formalism would be to show why elliptic Calabi-Yau are good solutions to the 12-dimensional theory. This would likely involve the construction of the supersymmetric version of the $\G$ EFT in order to study the generalised Killing spinor equation. 

Another interesting area of investigation would be the Heterotic/type II duality, where we should then consider EFT on a K3 background and relate this to Heterotic DFT \cite{Hohm:2011ex,Grana:2012rr}.

An interesting consequence of this work is that it shows how F-theory fits into a general picture of EFT with various $\Edd$ groups. One might be then be inspired to consider far more general backgrounds with higher dimensional fibres and with monodromies in $\Edd$ and so one would not just have sevenbranes but more exotic objects (of the type described in \cite{deBoer:2012ma}). In fact this idea appeared early in the F-theory literature \cite{Kumar:1996zx}. More recent work in this direction has appeared where one takes the fibre to be $K3$ and then one has a U-duality group act on the $K3$ \cite{Braun:2013yla,Candelas:2014jma,Candelas:2014kma} in a theory sometimes called G-theory. Further, the dimensional reduction of EFTs has now been examined in some detail and in particular one can make use of Scherk-Schwarz type reductions that yield gauged supergravities \cite{Aldazabal:2011nj,Geissbuhler:2011mx,Grana:2012rr,Berman:2012uy,Musaev:2013rq,Aldazabal:2013mya,Geissbuhler:2013uka,Berman:2013cli,Condeescu:2013yma,Aldazabal:2013via,Lee:2014mla,Baron:2014yua,Lee:2015xga}. This shows that one should perhaps consider more general type of reduction than the simple fibrations described here. This means one could consider Scherk-Schwarz type reductions on the F-theory torus. This makes no sense from the IIB perspective but it does from the point of view of the $\G$ EFT.

A futher quite radical notion would be the EFT version of a T-fold where we only have a local choice of section so that the space is not globally described by IIB or M-theory. One could have a monodromy such that as one goes round a one-cycle in nine dimensions and then flips between the IIB section and M-theory section. This would exchange a wrapped membrane in the M-theory section with a momentum mode IIB section just as a T-fold swaps a wrapped string with a momentum mode. 
Note, this is not part of the $\G$ duality group and thus is not a U-fold. This is simply because with two isometries one has a $\mathbb{Z}_2$ choice of section that one can then twist.

\section*{Acknowledgements} 
We have had useful communications with Charles Strickland-Constable, Malcolm Perry, Cumrun Vafa and Yi-Nan Wang. This work was initiated at the CERN-TH workshop on ``Duality Symmetries in String and M-theories'' and DSB was resident during the completion of the paper at the Erwin Schr\"odinger Institute ``Higher Structures in String Theory and Quantum Field Theory'' meeting. DSB is supported by the STFC grant ST/L000415/1 ``String Theory, Gauge Theory and Duality''. CB is supported in part by the Belgian Federal Science Policy Office through the Interuniversity Attraction Pole P7/37 ``Fundamental Interactions'', and in part by the ``FWO-Vlaanderen'' through the project G.0207.14N and by the Vrije Universiteit Brussel through the Strategic Research Program ``High-Energy Physics''. EM is supported by the ERC Advanced Grant ``Strings and Gravity" (Grant No. 32004). FJR is supported by an STFC studentship.

\appendix

\section{Cartan calculus, tensor hierarchy and the topological term}

The field content of exceptional field theories include in addition to the external metric and generalised metric a sequence of forms transforming under various representations of $\G$. These constitute the tensor hierarchy of EFT (similar to that of gauged supergravity \cite{deWit:2005hv,deWit:2008ta}). As well as being forms with respect to the ``external'' directions $x^\mu$, one may think of these fields as providing an analogue of forms from the point of view of the extended space. In this appendix we discuss the definitions and properties of these fields.

\subsection{Cartan calculus}
\label{sec:cartancalc}

Here we summarise the ``Cartan calculus'' relevant for the $\G$ EFT, discussed in \cite{Hohm:2015xna,Wang:2015hca}, in order to introduce our conventions. The form fields that we consider are valued in various representations, $\TAw$, $\TBw$, $\dots$ of $\G$. We list these in table \ref{t:Forms}. The value $w$ in brackets is the weight under generalised Lie derivatives. 

\vspace{1em}
\noindent\makebox[\textwidth]{
 \begin{minipage}{\textwidth}
  \begin{center}
  \begin{tabular}{|c|c|c|c|}
  \hline
   Module($w$) & Representations & Gauge field & Field strength \Tstrut\Bstrut \\ \hline
   $\TAw$ & $\mathbf{2}_{1}\oplus\mathbf{1}_{-1}$ & $\Aa^{\alpha} \oplus \Aa^{s}$ & $\Fa^{\alpha} \oplus \Fa^{s}$ \\
   $\TBw$ & $\mathbf{2}_{0}$ & $\Ab^{\alpha,s}$ & $\Fb^{\alpha,s}$ \\
   $\TCw$ & $\mathbf{1}_{1}$ & $\Ac^{\alpha\beta,s}$ & $\Fc^{\alpha\beta,s}$ \\
   $\TDw$ & $\mathbf{1}_{0}$ & $\Ad^{\alpha\beta,ss}$ & $\Fd^{\alpha\beta,ss}$ \\
   $\TEw$ & $\mathbf{2}_{1}$ & $\Ae^{\gamma,\alpha\beta,ss}$ & $\Fe^{\gamma,\alpha\beta,ss}$ \\
   $\TFw$ & $\mathbf{2}_{0}\oplus\mathbf{1}_{2}$ & $\Af_{\alpha} \oplus \Af_{s}$ & -- \\
   \hline
  \end{tabular}
 \vskip-0.5em
 \captionof{table}{\small{Modules, gauge fields and field strengths relevant for the tensor hierarchy and their representations under $\G$. The subscript denotes the weight under the $\mathbb{R}^+$. $w$ denotes the weights of the elements of the module under the generalised Lie derivative.}  } 
 \label{t:Forms}
  \end{center}
 \end{minipage}
}
\vspace{1em}

Given these representations, the key ingredients of the Cartan calculus are then a nilpotent derivative $\hpartial$ and an ``outer product'' $\p$ which act to map between the various modules listed in table \ref{t:Forms}. As first discussed in \cite{Cederwall:2013naa}, the chain complex
\begin{equation}
 \TAw \xleftarrow{~\hpartial~} \TBw \xleftarrow{~\hpartial~} \TCw \xleftarrow{~\hpartial~} \TDw \xleftarrow{~\hpartial~} \TEw \xleftarrow{~\hpartial~} \TFw \,,
\end{equation}
formed from these modules together with the nilpotent derivative can be seen as a generalisation of the deRham-complex and thus of differential forms. The bilinear product $\p$ is defined between certain modules such that it maps as follows:

\vspace{1em}
\noindent\makebox[\textwidth]{
 \begin{minipage}{\textwidth}
  \begin{center}
  \begin{tabular}{c| c c c ccc}
   $\bullet$ & $\TAw$ & $\TBw$ & $\TCw$ & $\TDw$ & $\TEw$ & $\TFw$ \\
   \hline
   $\TAw$ & $\TBw$ & $\TCw$ & $\TDw$ & $\TEw$ & $\TFw$ & $\TSw$ \\
   $\TBw$ & $\TCw$ & $\TDw$ & $\TEw$ & $\TFw$ & $\TSw$ & \\
   $\TCw$ & $\TDw$ & $\TEw$ & $\TFw$ & $\TSw$ & & \\
   $\TDw$ & $\TEw$ & $\TFw$ & $\TSw$ & & & \\
   $\TEw$ & $\TFw$ & $\TSw$ & & & & \\
   $\TFw$ & $\TSw$ & & & & & 
  \end{tabular}
  \end{center}
 \end{minipage}
}\\

\noindent The nilpotent derivative $\hpartial$ and the product $\p$ satisfy the following ``magic identity'' \cite{Hohm:2015xna,Wang:2015hca}
\begin{equation}
 \gL_\Lambda X = \Lambda \p \hpartial X + \hpartial \left( \Lambda \p X \right) \,,
\end{equation}
for all $X \in \TBw$, $\TCw$, $\TDw$, $\TEw$ and $\Lambda \in \TAw$. Here $\gL$ denotes the generalised Lie derivative \eqref{eq:gld}. Explicitly, the product is defined as
\begin{align}
  \left(\Aa_1 \p \Aa_2\right)^{\alpha,s} &= \Aa^\alpha_1 \Aa^s_2 + \Aa^s_1 \Aa^\alpha_2 \,, &  \left(\Aa \p \Ab\right)^{[\alpha\beta],s} &= 2 \Aa^{[\alpha} \Ab^{\beta],s} \,, \nonumber \\
  \left(\Aa \p \Ac\right)^{[\alpha\beta],ss} &= \Aa^s \Ac^{[\alpha\beta],s} \,, &  \left( \Aa \p \Ad \right)^{\gamma,[\alpha\beta],ss} &= \Aa^{\gamma} \Ad^{[\alpha\beta],ss} \,, \nonumber \\
  \left( \Aa \p \Ae \right)_{\gamma} &= \epsilon_{\gamma\delta} \Aa^s \Ae^{\gamma,[\alpha\beta],ss} \,, & \left( \Aa \p \Ae \right)_s &= \frac{1}{2}\epsilon_{\alpha\beta}\epsilon_{\gamma\delta} \Aa^{\gamma} E^{\delta,[\alpha\beta],ss} \,, \nonumber \\
  \left( \Aa \p \Af \right) &= \Aa^{\alpha} \Af_\alpha + \Aa^s \Af_s \,, & \left( \Ab_1 \p \Ab_2 \right)^{[\alpha\beta],ss} &= 2\Ab^{[\alpha|,s}_1 \Ab_2^{\beta],s} \,, \nonumber \\
  \left( \Ab \p \Ac \right)^{\gamma,[\alpha\beta],ss} &= \Ab^{\gamma,s} \Ac^{[\alpha\beta],s} \,, & \left( \Ab \p \Ad \right)_{\gamma} &= \frac{1}{2} \epsilon_{\alpha\beta} \epsilon_{\gamma\delta} \Ab^{\delta,s} \Ad^{[\alpha\beta],ss} \,, \nonumber \\
  \left( \Ab \p \Ad \right)_{s} &= 0 \,, & \left( \Ab \p \Ae \right) &= \frac{1}{2} \epsilon_{\alpha\beta} \epsilon_{\gamma\delta} \Ab^{\gamma,s} \Ae^{\delta,\alpha\beta,ss} \,, \nonumber \\
  \left( \Ac_1 \p \Ac_2 \right)_\gamma &= 0 \,, & \left( \Ac_1 \p \Ac_2 \right)_s &= \Ac_1^{[\alpha\beta],s} \Ac_2^{[\gamma\delta],s} \,, \nonumber \\
  \left( \Ac \p \Ad \right) &= \frac{1}{4} \epsilon_{\alpha\beta} \epsilon_{\gamma\delta} \Ac^{[\alpha\beta],s} \Ad^{[\gamma\delta],ss} \,, &  &
\end{align}
and is symmetric when acting on two different modules. The nilpotent derivative is defined as
\begin{align}
 \left( \hpartial \Ab \right)^{\alpha} &= \partial_s \Ab^{\alpha,s} \,, & \left( \hpartial \Ab \right)^{s} &= \partial_\alpha \Ab^{\alpha,s} \,, & \left( \hpartial \Ac \right)^{\alpha,s} &= \partial_\beta \Ac^{[\beta\alpha],s} \,, \nonumber \\
 \left( \hpartial \Ad \right)^{[\alpha\beta],s} &= \partial_s \Ad^{[\alpha\beta],ss} \,, & \left( \hpartial \Ae \right)^{[\alpha\beta],ss} &= \partial_\gamma \Ae^{\gamma,[\alpha\beta],ss} \,, & \left( \hpartial \Af \right)^{\gamma,[\alpha\beta],ss} &= \epsilon^{\gamma\delta} \partial_s F_\delta \,.
\end{align}
Nilpotency follows from the section condition \eqref{eq:constraint}.

Let us finally discuss some properties of the generalised Lie derivative which will be important in the construction of the tensor hierarchy in the next section. From now onwards we will often omit the $\G$ indices on elements of the modules. First, note that for any $\Aa_1$, $\Aa_2 \in \TAw$, the symmetric part of the Lie derivative is given by
\begin{equation}
 2 \left(\Aa_1 \,, \Aa_2 \right) \equiv \left( \gL_{\Aa_1} \Aa_2 + \gL_{\Aa_2} \Aa_1 \right) = \hpartial \left( \Aa_1 \p \Aa_2 \right) \,. \label{eq:SymBracket}
\end{equation}
Using the explicit formulae, one can see that this generates a vanishing generalised Lie derivative, i.e.
\begin{equation}
 \gL_{\left(\Aa_1 \,, \Aa_2\right)} = 0 \,.
\end{equation}
It will also be useful to write the generalised Lie derivative in terms of its symmetric and antisymmetric parts
\begin{equation}
 \gL_{\Aa_1} \Aa_2 = \left[ \Aa_1 \,, \Aa_2 \right]_E + \left( \Aa_1 \,, \Aa_2 \right) \,, \label{eq:gLsa}
\end{equation}
where the $E$-bracket is the antisymmetric part of the generalised Lie derivative
\begin{equation}
 \left[ \Aa_1 \,, \Aa_2 \right]_E = \frac12 \left( \gL_{\Aa_1} \Aa_2 - \gL_{\Aa_2} \Aa_1 \right) \,. \label{eq:EBracket}
\end{equation}
Finally, the Jacobiator of the $E$-bracket is proportional to terms that generate vanishing generalised Lie derivatives
\begin{equation}
 \left[ \left[ \Aa_1 \,, \Aa_2 \right]_E \,, \Aa_3 \right] + \textrm{cycl.} = \frac{1}{3} \left( \left[ \Aa_1 \,, \Aa_2 \right]_E \,, \Aa_3 \right) + \textrm{cycl.} \,, \label{eq:Jacobiator}
\end{equation}
so that the Jacobiator of generalised Lie derivatives does not vanish but lies in the kernel of the generalised Lie derivative when viewed as a map from generalised vectors to generalised tensors.

\subsection{Tensor hierarchy}
\label{sec:tensorhier}
The EFT is invariant under generalised diffeomorphisms, generated by a generalised vector field $\Lambda(x,Y) \in \TAw$. From the perspective of the ``extended space'' it induces gauge transformations and diffeomorphisms, while from the 9-dimensional perspective, it induces non-abelian gauge transformations of the scalar sector. Correspondingly, one introduces a gauge field $\Aa \in \TAw$ such that
\begin{equation}
 \delta_\Lambda \Aa = D_\mu \Lambda \,,
\end{equation}
where $D_\mu = \partial_\mu - \gL_{\Aa_\mu}$ is the 9-dimensional covariant derivative.\footnote{We will label gauge potentials $\Ad_{\mu_1\ldots\mu_4}$ in $\TDw$ by the same symbol but the 9-dimensional index can be used to distinguish between the gauge potential and the covariant derivative.} The naive form of the field strength would resemble the Yang-Mills field strength
\begin{equation}
 F_{\mu\nu} = 2 \partial_{[\mu} \Aa_{\nu]} - \left[ \Aa_\mu \,, \Aa_\nu \right]_E \,,
\end{equation}
which involves the $E$-bracket in order to be a 2-form. However, this fails to be gauge invariant
\begin{equation}
 \delta F_{\mu\nu} = 2 D_{[\mu} \delta \Aa_{\nu]} + \hpartial \left( \Aa_{[\mu} \p \delta \Aa_{\nu]} \right) \,,
\end{equation}
using \eqref{eq:gLsa}. In order to define a gauge-invariant field strength we are led to modify the usual Yang-Mills definition as follows
\begin{equation}
 \Fa_{\mu\nu} = 2 \partial_{[\mu} \Aa_{\nu]} - \left[ \Aa_{\mu}, \Aa_{\nu} \right]_E + \hpartial \Ab_{\mu\nu} \,,
\end{equation}
where $\Ab_{\mu\nu} \in \TBw$ is a 2-form. The modified field strength is now gauge-invariant if we define the variation of $\Ab_{\mu\nu}$ to be
\begin{equation}
 \Delta_\Lambda \Ab_{\mu\nu} = \Lambda \p \Fa_{\mu\nu} \,,
\end{equation}
where
\begin{equation}
 \Delta \Ab_{\mu\nu} \equiv \delta \Ab_{\mu\nu} + \Aa_{[\mu} \p \delta \Aa_{\nu]} \,.
\end{equation}
The definition of both the naive field strength $F_{\mu\nu}$ and the covariant field strength $\Fa_{\mu\nu}$ is compatible with the commutator of covariant derivatives
\begin{equation}
 \left[ D_{\mu} ,\, D_{\nu} \right] = - \gL_{F_{\mu\nu}} = - \gL_{\Fa_{\mu\nu}} \,,
\end{equation}
since their difference is of the form \eqref{eq:SymBracket} and thus generates vanishing generalised Lie derivative. Mirroring the tensor hierarchy of gauged supergravities \cite{deWit:2005hv,deWit:2008ta}, one can introduce a gauge transformation and field strength for $\Ab_{\mu\nu}$, which in turn requires a new 3-form potential. This way one obtains a hierarchy of $p$-form fields up to a 5-form gauge potential and its 6-form field strength. The 6-form potential does not appear in the action and so we do not define its field strength. Here we give this construction explicitly for the $\G$ EFT for the first time. In the following, the expressions for the 5-form potential and 6-form field strengths are new while the lower form potentials are also given in \cite{Hohm:2015xna}. Let us begin with the definition of the field strengths:
\begin{equation}
\begin{split}
\Fa_{\1\2} &= 2\partial_{[\1} \Aa_{\2]} - [\Aa_\1,\Aa_\2]_E + \hd\Ab_{\1\2} \,, \\
\Fb_{\1\2\3} &= 3\D_{[\1}\Ab_{\2\3]} - 3\partial_{[\1}\Aa_{\2}\pl\Aa_{\3]} + \Aa_{[\1}\pl[\Aa_\2,\Aa_{\3]}]_E + \hd\Ac_{\1\2\3} \,, \\
\Fc_{\1\ldots\4} &= 4\D_{[\1}\Ac_{\2\3\4]} + 3\hd \Ab_{[\1\2}\pl\Ab_{\3\4]} - 6\Fa_{[\1\2}\pl\Ab_{\3\4]} + 4\Aa_{[\1}\pl(\Aa_\2\pl\partial_\3\Aa_{\4]}) \\
 & \quad - \Aa_{[\1}\pl(\Aa_\2\pl[\Aa_\3,\Aa_{\4]}]_E) + \hd\Ad_{\1\2\3\4} \,, \\
\Fd_{\1\ldots\5} &= 5\D_{[\1}\Ad_{\2\ldots\5]} + 15\Ab_{[\1\2}\pl\D_3\Ab_{\4\5]} - 10\Fa_{[\1\2}\pl\Ac_{\3\4\5]} - 30\Ab_{[\1\2}\pl(\Aa_\3\pl\partial_\4\Aa_{\5]}) \\ 
& \quad + 10\Ab_{[\1\2}\pl(\Aa_\3\pl[\Aa_\4,\Aa_{\5]}]_E) - 5\Aa_{[\1}\pl(\Aa_\2\pl(\Aa_\3\pl\partial_\4\Aa_{\5]})) \\
 & \quad + \Aa_{[\1}\pl(\Aa_\2\pl(\Aa_\3\pl[\Aa_\4,\Aa_{\5]}]_E)) + \hd\Ae_{\1\ldots\5} \,, \\
\Fe_{\1\ldots\6} &= 6\D_{[\1}\Ae_{\2\ldots\6]} - 15\Fa_{[\1\2}\pl\Ad_{\3\ldots\6]} - 10\Ac_{[\1\2\3}\pl\hd\Ac_{\4\5\6]} - 20\Fb_{[\1\2\3}\pl\Ac_{\4\5\6]} \\
 & \quad - 45\Ab_{[\1\2}\pl(\hd\Ab_{\3\4}\pl\Ab_{\5\6}]) - 90\Ab_{[\1\2}\pl(\partial\Aa_{\3\4}\pl\Ab_{\5\6]})\\
		&\quad  + 45\Ab_{[\1\2}\pl([\Aa_\3,\Aa_\4]_E\,\pl\Ab_{\5\6}])  + 60\Ab_{[\1\2}\pl(\Aa_\3\pl(\Aa_\4\pl\partial_\5\Aa_{\6]})) \\
		& \quad - 15\Ab_{[\1\2}\pl(\Aa_\3\pl(\Aa_\4\pl[\Aa_\5,\Aa_{\6]}]_E))	+ 6\Aa_{[\1}\pl(\Aa_\2\pl(\Aa_\3\pl(\Aa_\4\pl\partial_\5\Aa_{\6]}))) \\
		& \quad -  \Aa_{[\1}\pl(\Aa_\2\pl(\Aa_\3\pl(\Aa_\4\pl[\Aa_\5,\Aa_{\6]}]_E))) + \hd\Af_{\1\ldots\6} 
\end{split}
\label{eq:fieldstrengths}
\end{equation}
Their variations are given by
\begin{equation}
\begin{split}
 \delta\Fa_{\1\2} &= 2\D_{[\1}\delta\Aa_{\2]} + \hd\Delta\Ab_{\1\2} \,, \\
 \delta\Fb_{\1\2\3} &= 3\D_{[\1}\Delta\Ab_{\2\3]} - 3\delta\Aa_{[\1}\pl\Fa_{\2\3]} + \hd\Delta\Ac_{\1\2\3} \,, \\
 \delta\Fc_{\1\ldots\4} &= 4\D_{[\1}\Delta\Ac_{\2\3\4]} - 4\delta\Aa_{[\1}\pl\Fb_{\2\3\4]} - 6\Fa_{[\1}\pl\Delta\Ab_{\2\3\4]} + \hd\Delta\Ad_{\1\ldots\4} \,, \\
 \delta\Fd_{\1\ldots\5} &= 5\D_{[\1}\Delta\Ad_{\2\ldots\5]} - 5\delta\Aa_{[\1}\pl\Fc_{\2\ldots\5]} - 10\Fa_{[\1\2}\pl\Delta\Ac_{\3\4\5]} - 10\Fb_{[\1\2\3}\pl\Delta\Ab_{\4\5]} \\
 & \quad + \hd\Delta\Ae_{\1\ldots\5} \,, \\
 \delta\Fe_{\1\ldots\6} &= 6\D_{[\1}\Delta\Ae_{\2\ldots\6]} - 6\delta\Aa_{[\1}\pl\Fd_{\2\ldots\6]} - 15\Fa_{[\1\2}\pl\Delta\Ad_{\3\ldots\6]} - 20\Fb_{[\1\2\3}\pl\Delta\Ac_{\4\5\6]} \\
 & \quad + 15\Fc_{[\1\ldots\4}\pl\Delta\Ab_{\5\6]} + \hd\Delta\Af_{\1\ldots\6} \,,
\end{split}
\label{eq:fieldstrengthvariation}
\end{equation}
where have we defined the ``covariant'' gauge field variations as
\begin{equation}
 \begin{split}
 \Delta\Ab_{\1\2} &= \delta\Ab_{\1\2} + \Aa_{[\1}\pl\delta\Aa_{\2]} \,, \\
 \Delta\Ac_{\1\2\3} &= \delta\Ac_{\1\2\3} - 3\delta\Aa_{[\1}\pl\Ab_{\2\3]} + \Aa_{[\1}\pl(\Aa_\2\pl\delta\Aa_{\3]}) \,, \\
 \Delta\Ad_{\1\ldots\4} &= \delta\Ad_{\1\ldots\4} - 4\delta\Aa_{[\1}\pl\Ac_{\2\ldots\4]} + 3\Ab_{[\1\2}\pl\left(\delta\Ab_{\3\4]}+2\Aa_{\3}\pl\delta\Aa_{\4]}\right) + \Aa_{[\1}\!\pl(\Aa_\2\pl(\Aa_\3\pl\delta\Aa_{\4]})) \,, \\
 \Delta\Ae_{\1\ldots\5} &= \delta\Ae_{\1\ldots\5} - 5\delta\Aa_{[\1}\pl\Ad_{\2\ldots\5]} - 10\delta\Ab_{[\1\2}\pl\Ac_{\3\ldots\5]} - 15\Ab_{[\1\2}\pl\left(\delta\Aa_\3\pl\Ab_{\4\5]}\right) \\
 & \quad -10\left(\Aa_{[\1}\pl\delta\Aa_\2\right)\pl\Ac_{\3\4\5]} + 10\Ab_{[\1\2}\pl\left(\Aa_\3\pl\left(\Aa_\4\pl\delta\Aa_\5\right)\right) + \Aa_{[\1}\!\pl\left(\Aa_\2\pl\left(\Aa_\3\pl\left(\Aa_\4\pl\delta\Aa_{\5]}\right)\right)\right) \,, \\
 \Delta\Af_{\1\ldots\6} &= \delta\Af_{\1\ldots\6} - 6\delta\Aa_{[\1}\pl\Ae_{\2\ldots\6]} - 15\delta\Ab_{[\1\2}\pl\Ad_{\3\ldots\6]} - 15\left(\Aa_{[\1}\pl\delta\Aa_\2\right)\,\pl\Ad_{\3\ldots\6]} \\
 &\quad	- 10\delta\Ac_{[\1\2\3}\pl\Ac_{\4\5\6]} + 60(\delta\Aa_{[\1}\pl\Ab_{\2\3}\pl\Ac_{\4\5\6]} - 20\left(\Aa_{[\1}\pl\left(\Aa_\2\pl\delta\Aa_\3\right)\right)\,\pl\Ac_{\4\5\6]} \\
 &\quad	- 45 \Ab_{[\1\2}\pl\left(\delta\Ab_{\3\4}\pl\Ab_{\5\6]}\right) + 45 \Ab_{[\1\2}\pl\left(\Ab_{\3\4}\pl(\Aa_\5\pl\delta\Aa_{\6]}\right)  \\
 &\quad	+ 15 \Ab_{[\1\2}\pl\left(\Aa_\3\pl\left(\Aa_\4\pl\left(\Aa_\5\pl\delta\Aa_{\6]}\right)\right)\right) + \Aa_{[\1}\pl\left(\Aa_\2\pl\left(\Aa_\3\pl\left(\Aa_\4\pl\left(\Aa_\5\pl\delta\Aa_{\6]}\right)\right)\right)\right) \,.
 \end{split}
 \label{eq:gaugefieldvariation}
\end{equation}
It is now easy to check that the field strengths \eqref{eq:fieldstrengths} are invariant under the following gauge transformations
\begin{equation}
 \begin{split}
  \delta \Aa_{\1} &= D_{\1} \Lambda - \hat{\partial}\, \Xi_\1 \,, \\
  \Delta \Ab_{\1\2} &= \Lambda\, \pl \Fa_{\1\2} + 2 D_{[\1} \Xi_{\2]} - \hat{\partial} \Theta_{\1\2} \,, \\
  \Delta \Ac_{\1\ldots\3} &= \Lambda\, \pl \Fb_{\1\2\3} + 3 \Fa_{[\1\2} \pl \Xi_{\3]} + 3 D_{[\1} \Theta_{\2\3]} - \hat{\partial} \Omega _{\1\ldots\3} \,, \\
  \Delta \Ad_{\1\ldots\4} &= \Lambda\, \pl \Fc_{\1\ldots \4} - 4 \Fb_{[\1\ldots\3} \pl \Xi_{\4]} + 6 \Fa_{[\1\2} \pl \Theta_{\3\4]} + 4 D_{[\1} \Omega_{\2\ldots\4]} - \hat{\partial} \Upsilon_{\1\ldots\4} \,, \\
  \Delta \Ae_{\1\ldots\5} &= \Lambda\, \pl \Fd_{\1\ldots\5} - 5 \Fc_{[\1\ldots\4} \pl \Xi_{\5]} - 10 \Fb_{[\1\ldots\3} \pl \Theta_{\4\5]} + 10 \Fa_{[\1\2} \pl \Omega_{\3\ldots \5]} \\
  & \quad + 5 D_{[\1} \Upsilon_{\2 \ldots \5]} - \hat{\partial} \Phi_{\1\ldots\5} \,, \\
  \Delta \Af_{\1 \ldots \6} &= \Lambda\, \pl {\cal L}_{\1\ldots \6} + 6 \Fd_{[\1\ldots\5} \pl \Xi_{\6]} + 15 \Fc_{\1\ldots\4} \pl \Theta_{\5\6]} - 20 \Fb_{[\1\ldots\3} \pl \Omega_{\4\ldots\6]} \\
  & \quad + \Fa_{[\1\2} \pl \Upsilon_{\3\ldots\6]} + 6 D_{[\1} \Phi_{\2 \ldots \6]} \,.
 \end{split}
\end{equation}
Finally, the field strengths \eqref{eq:fieldstrengths} satisfy the following Bianchi identities, as can be seen from their definitions.
\begin{equation}
\begin{split}
3\D_{[\1}\Fa_{\2\3]} &= \hd\Fb_{\1\ldots\3} \,, \\
4\D_{[\1}\Fb_{\2\ldots\4]} + 3\Fa_{[\1\2}\pl\Fa_{\3\4]} &= \hd\Fc_{\1\ldots\4} \,, \\
5\D_{[\1}\Fc_{\2\ldots\5]} + 10\Fa_{[\1\2}\pl\Fb_{\3\4\5]} &= \hd\Fd_{\1\ldots\5} \,, \\
6\D_{[\1}\Fd_{\2\ldots\6]} + 15\Fa_{[\1\2}\pl\Fc_{\3\ldots\6]} -10\Fb_{[\1\2\3}\pl\Fb_{\4\5\6]} &= \hd\Fe_{\1\ldots\6} \,.
\end{split}
\label{eq:Bianchi}
\end{equation}
While the first three equations have appeared before, the final identity is new.

\subsection{Topological term}
\label{sec:topterm}
Maximal supergravity theories contain a topological term, which is mirrored in the corresponding EFT. Armed with the Cartan calculus and the tensor hierarchy we can now construct a topological term for the action. It is given by
\begin{equation}
 \begin{split}
  S_{top} &= \kappa \int \dd^{10}x\, \dd^3Y\, \varepsilon^{\1\ldots\mt} \left[ \frac{1}{5} \hpartial \Fd_{\1\ldots\5} \p \Fd_{\6\ldots\mt} - \frac{5}{2} \left( \Fa_{\1\2} \p \Fc_{\3\ldots\6} \right) \p \Fc_{\7\ldots\mt} \right. \\
  & \quad \left. 
+ \frac{10}{3} \left( \Fb_{\1\ldots\3} \p \Fb_{\4\ldots\6} \right) \p \Fc_{\7 \ldots \mt} \right] \,,
 \end{split} \label{eq:ToptTermApp}
\end{equation}
where we have abused the notation to also denote the 10-dimensional indices by $\1, \ldots, \mt = 1, \ldots, 10$, and $\varepsilon^{\1\ldots\mt} = \pm 1$ is the 10-dimensional alternating symbol. This term is a manifestly gauge-invariant boundary term in ten dimensions and has weight one under generalised diffeomorphisms, as required. Instead of explicitly showing that it is a boundary term itself, we will just show that its variation is a boundary term. Using the variations of the field strengths \eqref{eq:fieldstrengthvariation} and the Bianchi identities \eqref{eq:Bianchi}, one finds
\begin{equation}
 \begin{split}
  \delta S_{top} &= \kappa \int\!\! \dd^{10}x\, \dd^3Y\, \varepsilon^{\1\ldots\mt} D_{\1} 
  		\left[\vphantom{\frac{40}{3}} 	- 5 \left( \delta \Aa_{\2} \bullet {\cal J}_{\3 \ldots \6} \right) \bullet {\cal J}_{\7 \ldots \mt} \right. \\
  & \quad \left. 	+ 20 \left( \Delta \Ab_{\2\3} \bullet {\cal H}_{\4\ldots\6} \right) \bullet {\cal J}_{\7\ldots \mt}  
  		- 20 \left( {\cal F}_{\2\3} \bullet {\cal J}_{\4 \ldots \7} \right) \bullet \Delta \Ac_{\8 \ldots \mt} \right. \\
  & \quad \left. 	+ \frac{40}{3} \left( {\cal H}_{\2\ldots\4} \bullet {\cal H}_{\5\ldots\7} \right) \bullet \Delta \Ac_{\8\ldots\mt}  
  		+ 2 \hat{\partial} \Delta \Ad_{\2\ldots\5} \bullet {\cal K}_{\6 \ldots \mt} \right] \,. \label{eq:TopVariation}
 \end{split}
\end{equation}
Throughout we assume that the ``extended space'' parametrised by the $Y$'s does not have a boundary. One can easily check using \eqref{eq:gaugefieldvariation} that the term is invariant under gauge transformations. In the next appendix, we will use \eqref{eq:TopVariation} to fix the overall coefficient relative to the other terms in the action by requiring invariance under external diffeomorphisms.

\section{Determining the action}
\label{sec:detact}

\subsection{Anomalous variations}

We need to be able to cancel all ``anomalous variations'' under internal and generalised diffeomorphisms. These are defined to be the non-covariant parts of the transformation of some possibly non-tensorial object $\mathcal{T}$:
\be
\anom_\Lambda \mathcal{T} = \delta_\Lambda \mathcal{T} - \mathcal{L}_\Lambda \mathcal{T} \,,
\ee
for generalised diffeomorphisms, and 
\be
\anom_\xi \mathcal{T} = \delta_\xi \mathcal{T} - L_\xi \mathcal{T} \,,
\ee
under external diffeomorphisms. 
This can be applied to products and obeys a Leibniz-like property. 

\subsubsection{Generalised diffeomorphisms} 

In particular, we build our action using derivatives of the external and generalised metrics. Under generalised diffeomorphisms, one has
\be
\begin{split} 
\delta_\Lambda \mathcal{M}_{ss} & = \Lambda^M \partial_M \mathcal{M}_{ss} 
- \frac{12}{7} \partial_\alpha \Lambda^\alpha \mathcal{M}_{ss} 
+ \frac{16}{7} \partial_s \Lambda^s \mathcal{M}_{ss} \,, \\
\delta_\Lambda \mathcal{H}_{\alpha \beta} & = \Lambda^M \partial_M\mathcal{H}_{\alpha \beta}
+ \partial_\alpha \Lambda^\gamma \cH_{\gamma \beta} + \partial_\beta \Lambda^\gamma \cH_{\alpha \gamma} 
- \partial_\alpha \Lambda^\alpha \mathcal{H}_{\alpha \beta} 
\\ 
\delta_\Lambda g_{\mu \nu} & = \Lambda^M \partial_M g_{\mu \nu} + \frac{2}{7} \partial_M \Lambda^M g_{\mu \nu} \,.
\end{split}
\ee
The anomalous variations of the derivatives of these objects are
\be
\begin{split}
\anom_\Lambda \partial_s \gM_{ss}  & = \frac{16}{7} \partial_s\partial_s \Lambda^s \gM_{ss} \,, \\
\anom_\Lambda \partial_s \cH_{\alpha \beta} & = 0 \,, \\
\anom_\Lambda \partial_s g_{\mu \nu}  & = \frac{2}{7} \partial_s \partial_s \Lambda^s g_{\mu \nu} \,,
\end{split} 
\label{eq:anomvars} 
\ee
and
\be
\begin{split}
\anom_\Lambda \partial_\gamma \gM_{ss}  & = -\frac{12}{7} \partial_\gamma \partial_\beta \Lambda^\beta \gM_{ss} \,, \\
\anom_\Lambda \partial_\gamma \cH_{\alpha \beta} & = \partial_\gamma \partial_\alpha \Lambda^\delta \cH_{\delta \beta}
+ \partial_\gamma \partial_\alpha \Lambda^\delta \cH_{\alpha \delta} - \partial_\delta \partial_\gamma \Lambda^\delta \cH_{\alpha \beta}
  \,, \\
\anom_\Lambda \partial_\gamma g_{\mu \nu}  & = \frac{2}{7} \partial_\gamma \partial_\beta \Lambda^\beta g_{\mu \nu} \,.\\
\end{split} 
\ee
Note that one finds that it is necessary to treat the derivatives $\partial_M$ as carrying the special weight $-1/7$, as is a familiar aspect of the exceptional field theory construction \cite{Hohm:2013vpa}.

\subsubsection{External diffeomorphisms} 

Recall that the external derivative is written in a form covariant under generalised diffeomorphisms, using $D_\mu = \partial_\mu - \delta_{A_\mu}$. The variations of the fields under external diffeomorphisms are then
\be
\begin{split}
\delta_\xi g_{\mu \nu} 
 & = \xi^\rho D_\rho g_{\mu\nu} 
+ D_\mu \xi^\rho g_{\rho \nu} 
+ D_\nu \xi^\rho g_{\rho \mu} \,,\\
\delta_\xi \gM_{MN} & = \xi^\mu D_\mu \gM_{MN} \,, \\
\delta_\xi A_\mu{}^M & = \xi^\nu \mathcal{F}_{\nu \mu}{}^M + \gM^{MN} g_{\mu \nu} \partial_N \xi^\nu \,, \\ 
\Delta_\xi B_{\mu \nu} & = \xi^\lambda \mathcal{H}_{\lambda \mu \nu}  \,,\\ 
\Delta_\xi C_{\mu \nu \sigma} & = \xi^\lambda \mathcal{J}_{\lambda \mu \nu \sigma} \,,\\ 
\Delta_\xi D_{\mu \nu \sigma \kappa} & = \xi^\lambda \mathcal{K}_{\lambda \mu \nu \sigma \kappa} \,, \\ 
\Delta_\xi E_{\mu \nu \sigma \kappa \tau} & = \xi^\lambda \mathcal{L}_{\lambda \mu \nu \sigma \kappa \tau} \,,
\end{split}
\ee
where we define the variation of the higher rank gauge fields through their covariant variation $\Delta$, not to be confused with the anomalous variation $\anom$.

It is important to be careful as the derivatives $D_\mu$ do not commute. One has 
\be
[ D_\mu, D_\nu ] \xi^\rho = - \mathcal{F}_{\mu \nu}{}^M \partial_M \xi^\rho \,,
\ee
for a diffeomorphism gauge parameter $\xi^\mu$ which is a scalar under generalised diffeomorphisms. On the metric, which is weighted under generalised diffeomorphisms, one has 
\be
[ D_\mu , D_\sigma ] g_{\nu \rho} =
 - \mathcal{F}_{\mu \sigma}{}^M \partial_M g_{\nu \rho} 
 + 2 \omega \partial_M \mathcal{F}_{\mu \sigma}{}^M g_{\nu \rho} 
\ee
One must use these relationships for instance in reexpressing the bare variation $\delta_\xi D_\mu g_{\nu \rho}$ in terms of the Lie derivative $L_\xi D_\mu g_{\nu \rho}$. One finds that
\be
\begin{split} 
\delta_\xi D_\mu g_{\nu \rho} 
& = 
D_\mu \delta_\xi g_{\nu \rho} - \delta_{(\delta_\xi A_\mu)} g_{\nu \rho} 
\\ & = L_\xi ( D_\mu g_{\nu \rho} ) 
+ D_\mu D_\nu \xi^\sigma g_{\sigma \rho} 
+ D_\mu D_\rho \xi^\sigma g_{\nu \sigma} 
- \MM^{MN} g_{\mu \sigma} \partial_N \xi^\sigma \partial_M g_{\nu \rho} \\ 
& \qquad
+ 2 \omega \mathcal{F}_{\sigma \mu}{}^M \partial_M \xi^\sigma  g_{\nu \rho} 
+ 2 \omega \partial_M \left( \MM^{MN} g_{\mu \sigma} \partial_N \xi^\sigma \right) g_{\nu \rho} \,.
\end{split}
\ee
Similar treatment of derivatives $D_\mu$ of the scalars gives
\be
\begin{split}
\anom_\xi D_\mu \ln \MM_{ss} & = 
\frac{12}{7} \partial_\gamma \xi^\nu \mathcal{F}_{\nu \mu}{}^\gamma 
- \frac{16}{7} \partial_s \xi^\nu \mathcal{F}_{\nu \mu}{}^s
-\MM^{MN} g_{\mu \nu} \partial_N \xi^\nu \partial_M \ln \MM_{ss} \\ 
& \qquad + \frac{12}{7} \partial_\gamma \left( \MM^{\gamma \beta} g_{\mu \nu} \partial_\beta \xi^\nu \right) 
- \frac{16}{7} \partial_s\Big( \MM^{ss} g_{\mu \nu} \partial_s \xi^\nu \Big) 
\end{split} 
\ee
and
\be
\begin{split}
\anom_\xi D_\mu \cH_{\alpha \beta} & = 
- \MM^{MN} g_{\mu \nu} \partial_N \xi^\nu \partial_M \cH_{\alpha \beta} 
\\ 
& \qquad
- \partial_\alpha \xi^\nu \mathcal{F}_{\nu \mu}{}^\gamma \cH_{\gamma \beta} 
- \partial_\beta \xi^\nu \mathcal{F}_{\nu \mu}{}^\gamma \cH_{\gamma \alpha} 
- \partial_\gamma \xi^\nu \mathcal{F}_{\nu \mu}{}^\gamma \cH_{\alpha\beta} \\ 
& \qquad
- \partial_\alpha \left( \MM^{\gamma \delta} g_{\mu \nu} \partial_\delta \xi^\nu \right) \cH_{\gamma \beta} 
- \partial_\beta \left( \MM^{\gamma \delta} g_{\mu \nu} \partial_\delta \xi^\nu \right) \cH_{\alpha \gamma} 
- \partial_\gamma \left( \MM^{\gamma \delta} g_{\mu \nu} \partial_\delta \xi^\nu \right) \cH_{\alpha \beta} 
\end{split} 
\ee
Finally, we should consider the variation under external diffeomorphisms of the derivatives of tensors with respect to the internal coordinate. One finds here simply
\be
\begin{split} 
\anom_\xi \partial_P  \gM_{MN} & = \partial_P \xi^\mu D_\mu  \gM_{MN}  \,, \\
\Delta_\xi \partial_M g_{\mu \nu} & = \partial_M \xi^\rho D_\rho g_{\mu \nu}
 + D_\mu \partial_M \xi^\rho g_{\rho \nu} 
 + D_\nu \partial_M \xi^\rho g_{\rho \mu}  \,.\\
\end{split}
\ee

\subsubsection{Variations of field strengths}

Using the transformations of the field strengths under arbitrary variations of the gauge fields \eqref{eq:fieldstrengthvariation}, one finds that under external diffeomorphisms
\be
\begin{split} 
\anom_\xi \mathcal{F}_{\mu \nu}{}^\alpha & = \partial_s \xi^\rho \mathcal{H}_{\nu \mu \rho}{}^{\alpha s} + 2 D_{[\mu} \left( \gM^{\alpha \beta} \partial_\beta \xi^\lambda g_{\nu] \lambda} \right) \,,\\ 
\anom_\xi \mathcal{F}_{\mu \nu}{}^s & = \partial_\alpha \xi^\rho \mathcal{H}_{\nu \mu \rho}{}^{\alpha s} + 2 D_{[\mu} \left( \gM^{ss} \partial_s \xi^\lambda g_{\nu] \lambda} \right) \,, \\ 
\anom_\xi \mathcal{H}_{\mu \nu \rho}{}^{\alpha s} & = 
\partial_\beta \xi^\lambda J_{\sigma \mu \nu \rho}{}^{[\beta \alpha ] s} 
- 3 \MM^{\alpha \beta} \partial_\beta \xi^\lambda g_{\lambda[\mu} \mathcal{F}_{\nu \rho]}{}^s 
- 3 \MM^{ss} \partial_s \xi^\lambda g_{\lambda[\mu} \mathcal{F}_{\nu \rho]}{}^\alpha \,, \\ 
\anom_\xi \mathcal{J}_{\mu \nu \rho \sigma}{}^{[\alpha \beta] s} & = \partial_s \xi^\lambda \mathcal{K}_{\lambda \mu \nu \rho \sigma}{}^{[\alpha \beta]ss} 
- 8 \MM^{\gamma [ \alpha } \partial_\gamma \xi^\lambda g_{\lambda[\mu} \mathcal{H}_{\nu \rho \sigma]}{}^{\beta]s} \,.
\end{split} 
\ee
(By construction of course the field strengths are invariant under generalised diffeomorphisms which manifest as the gauge transformation of $A_\mu{}^M$.)

\subsection{Varying the action}

The action can now be determined using what is by now a fairly standard procedure. One writes down the general possible form of a two-derivative scalar Lagrangian, ensuring that all terms carry the correct weights under the diffeomorphism symmetries (and also under the $\mathbb{R}^+$ factor of the duality group). Then one varies this under the above local symmetries and chooses coefficients such that the anomalous variations, as defined in the previous subsections, cancel. This turns out to fix the Lagrangian uniquely, up to overall scale and some total derivatives. 

This calculation is straightforward but somewhat tedious to relate. For the reader familiar already with the exceptional field theory construction no particularly interesting technical novelties arise; while the reader unfamiliar with the field is perhaps better encouraged to believe the correctness of the result and may find it more illuminating to study the final form of the theory as given in the bulk of the paper. Nevertheless, we shall give a sketch of the details. 

For instance, let us consider only terms which involve the derivative $\partial_s$ (that we can do this in the first place is an unusual consequence of the relative simplicity of the $\G$ coordinate representation and section condition). The most general ansatz for a ``potential term'' which is a scalar under generalised diffeomorphisms up to section condition is
\be
\begin{split} 
V_1 &  = a_0 \gM^{ss} \partial_s \cH^{\alpha \beta} \partial_s \cH_{\alpha \beta} 
 + a_1 \gM^{ss} \partial_s \gM^{ss} \partial_s \gM_{ss}
+ a_2 \partial_s \partial_s \gM^{ss} 
 + a_3 \gM^{ss} \partial_s g^{\mu \nu} \partial_s g_{\mu \nu} \\ & \qquad
+ a_4 \gM^{ss} \partial_s \ln g \partial_s \ln g
+ a_5 \gM^{ss} \partial_s \ln g \partial_s \gM_{ss} \gM^{ss} 
+ a_6 \gM^{ss} \partial_s \partial_s \ln g \,.
\end{split} 
\ee
One firstly attempts to fix this by varying under generalised diffeomorphisms. It is enough to calculate the anomalous variation using \eqref{eq:anomvars}. This produces the following constraints on the unknown coefficients:
\be
\frac{18}{7} a_6 = \frac{16}{7} a_2 \quad , \quad
a_6 - \frac{4}{7} a_3 + \frac{36}{7} a_4 + \frac{16}{7} a_5 =0 
\quad , \quad 
-\frac{25}{7} a_2 + \frac{32}{7} a_1 - \frac{18}{7} a_5 = 0 \,.
\ee
These are not sufficient to fully determine the form of $V_1$, even up to scale and total derivatives. Indeed, as $\partial_s \cH_{\alpha \beta}$ has no anomalous variation the coefficient of the single term involving its square cannot be determined here. 

One fixes the coefficients fully by varying under external diffeomorphisms. In doing so, it becomes especially convenient to drop all total derivatives. 

The anomalous variation of the scalar potential can then be made to cancel by incorporating contributions from that of the kinetic terms,
\be
\mathcal{L}_{skin} = d_1 g^{\mu \nu} D_\mu \ln \gM_{ss} D_\nu \ln \gM_{ss} 
+ d_2 g^{\mu \nu} D_\mu \cH_{\alpha \beta} D^\nu \cH^{\alpha \beta} \,,
\ee
and Ricci scalar, which up to total derivatives has the following anomalous variation which in fact is common to all exceptional field theories regardless of dimensions,
\be
\begin{split}
\anom_\xi \hat{R} & =   
\gM^{MN} \partial_N \xi^\lambda \Big( 
\partial_M D_\lambda \ln g - g^{\mu \nu} \partial_M D_\nu g_{\mu \lambda} \\ & \qquad
- \frac{1}{2} \partial_M g_{\nu \rho} D_\lambda g^{\nu \rho} 
- \frac{1}{2} g^{\mu \nu} D_\nu \ln g \partial_M g_{\mu \lambda} 
- \partial_M g_{\mu \lambda} D_\nu g^{\mu \nu} 
\Big) 
\\ & \qquad
- g^{\mu\nu} g^{\rho \sigma} D_\mu g_{\sigma \lambda} \mathcal{F}_{\rho \nu}{}^M \partial_M  \xi^\lambda 
- g^{\mu \nu} \mathcal{F}_{\rho \mu}{}^M \partial_M D_\nu \xi^\rho  \, .
\label{eq:RextFvar1}
\end{split}
\ee
In particular, ignoring for now any terms involving the field strength $\mathcal{F}_{\mu \nu}{}^M$,  one finds anomalous variations which are non-linear in the components of the generalised metric, namely
\be
2( a_0 -  d_2 ) \gM^{ss} D_\mu \cH^{\alpha \beta} \partial_s \cH_{\alpha \beta} \partial_s \xi^\mu  +
2 \left( a_1 -  \frac{9 d_1}{7} \right) \gM^{ss} \partial_s \gM^{ss} D_\mu \gM_{ss}   \partial_s \xi^\mu   \,.
\ee
This has the effect of fixing the coefficients of the scalar kinetic term relative to those appearing in $V_1$. 
In addition there are terms involving two derivatives of the external metric, which arise from the Ricci scalar, kinetic terms and scalar potentials, which cancel if 
\be
4 a_3 = 1 \quad , \quad
2 a_6 - 4 a_4 = -1 \quad , \quad
2a_5 + 2a_6 + a_2 = \frac{32d_1}{7} \,,
\ee
while additional terms involving derivatives of $\gM_{ss}$ cancel if 
\be
 4 a_3 = - \frac{32d_1}{7} \quad , \quad - \frac{1}{2} a_2 - 4 a_4  - a_5 + a_6 = \frac{16}{7} d_1 \,,
\ee
and finally one has a number of other terms which cancel if $a_3 = 1/4$, which is already required. 

One can similarly determine the form of $V_2$, which involves the derivatives $\partial_\alpha$ only. Altogether, the constraints that one obtains entirely fix the form of $\mathcal{L}_{skin}$ and the scalar potential $V$ up to possible total derivatives in $\partial_s, \partial_\alpha$, which can be ignored.

Next, one can consider the anomalous variations arising from the Ricci scalar and scalar kinetic terms which involve the field strength $\mathcal{F}_{\mu \nu}{}^M$. These can be arranged to cancel against part of the variation of the kinetic term $-\frac{1}{4} \mathcal{F}_{\mu \nu}{}^M \mathcal{F}^{\mu \nu N} \gM_{MN}$, fixing the relative coefficient of this term with respect to the the part of the action involving metric derivatives. 

At this point the cancellation of all terms involving metric derivatives and $\mathcal{F}_{\mu \nu}{}^M$ has been ensured. The remaining anomalous variations involve the variation of the gauge field kinetic terms and the topological term. With the coefficients shown in this paper, these can be easily demonstrated to cancel upon making use of the duality relation \eqref{eq:Deom} 
following from the equation of motion of $D_{\mu \nu \rho \sigma}$.

\bibliography{NewBib}

\providecommand{\href}[2]{#2}\begingroup\raggedright\begin{thebibliography}{10}

\bibitem{Witten:1995ex}
E.~Witten, {\it {String theory dynamics in various dimensions}},  {\em
  Nucl.Phys.} {\bf B443} (1995) 85--126,
  [\href{http://arxiv.org/abs/hep-th/9503124}{{\tt hep-th/9503124}}].

\bibitem{Hull:1994ys}
C.~Hull and P.~Townsend, {\it {Unity of superstring dualities}},  {\em
  Nucl.Phys.} {\bf B438} (1995) 109--137,
  [\href{http://arxiv.org/abs/hep-th/9410167}{{\tt hep-th/9410167}}].

\bibitem{Vafa:1996xn}
C.~Vafa, {\it {Evidence for F theory}},  {\em Nucl.Phys.} {\bf B469} (1996)
  403--418, [\href{http://arxiv.org/abs/hep-th/9602022}{{\tt hep-th/9602022}}].

\bibitem{Morrison:1996na}
D.~R. Morrison and C.~Vafa, {\it {Compactifications of F theory on Calabi-Yau
  threefolds. 1}},  {\em Nucl.Phys.} {\bf B473} (1996) 74--92,
  [\href{http://arxiv.org/abs/hep-th/9602114}{{\tt hep-th/9602114}}].

\bibitem{Morrison:1996pp}
D.~R. Morrison and C.~Vafa, {\it {Compactifications of F theory on Calabi-Yau
  threefolds. 2.}},  {\em Nucl.Phys.} {\bf B476} (1996) 437--469,
  [\href{http://arxiv.org/abs/hep-th/9603161}{{\tt hep-th/9603161}}].

\bibitem{Kumar:1996zx}
A.~Kumar and C.~Vafa, {\it {U manifolds}},  {\em Phys.Lett.} {\bf B396} (1997)
  85--90, [\href{http://arxiv.org/abs/hep-th/9611007}{{\tt hep-th/9611007}}].

\bibitem{Coimbra:2011ky}
A.~Coimbra, C.~Strickland-Constable, and D.~Waldram, {\it {$E_{d(d)} \times
  \mathbb{R}^+$ generalised geometry, connections and M theory}},  {\em JHEP}
  {\bf 1402} (2014) 054, [\href{http://arxiv.org/abs/1112.3989}{{\tt
  arXiv:1112.3989}}].

\bibitem{Coimbra:2012af}
A.~Coimbra, C.~Strickland-Constable, and D.~Waldram, {\it {Supergravity as
  Generalised Geometry II: $E_{d(d)} \times \mathbb{R}^+$ and M theory}},  {\em
  JHEP} {\bf 1403} (2014) 019, [\href{http://arxiv.org/abs/1212.1586}{{\tt
  arXiv:1212.1586}}].

\bibitem{Hitchin:2004ut}
N.~Hitchin, {\it {Generalized Calabi-Yau manifolds}},  {\em Quart.J.Math.Oxford
  Ser.} {\bf 54} (2003) 281--308,
  [\href{http://arxiv.org/abs/math/0209099}{{\tt math/0209099}}].

\bibitem{Gualtieri:2003dx}
M.~Gualtieri, {\it {Generalized complex geometry}},
  \href{http://arxiv.org/abs/math/0401221}{{\tt math/0401221}}.

\bibitem{Hull:2009mi}
C.~Hull and B.~Zwiebach, {\it {Double Field Theory}},  {\em JHEP} {\bf 0909}
  (2009) 099, [\href{http://arxiv.org/abs/0904.4664}{{\tt arXiv:0904.4664}}].

\bibitem{Hull:2009zb}
C.~Hull and B.~Zwiebach, {\it {The Gauge algebra of double field theory and
  Courant brackets}},  {\em JHEP} {\bf 0909} (2009) 090,
  [\href{http://arxiv.org/abs/0908.1792}{{\tt arXiv:0908.1792}}].

\bibitem{Hohm:2010jy}
O.~Hohm, C.~Hull, and B.~Zwiebach, {\it {Background independent action for
  double field theory}},  {\em JHEP} {\bf 1007} (2010) 016,
  [\href{http://arxiv.org/abs/1003.5027}{{\tt arXiv:1003.5027}}].

\bibitem{Hohm:2010pp}
O.~Hohm, C.~Hull, and B.~Zwiebach, {\it {Generalized metric formulation of
  double field theory}},  {\em JHEP} {\bf 1008} (2010) 008,
  [\href{http://arxiv.org/abs/1006.4823}{{\tt arXiv:1006.4823}}].

\bibitem{Duff:1989tf}
M.~Duff, {\it {Duality rotations in string theory}},  {\em Nucl.Phys.} {\bf
  B335} (1990) 610.

\bibitem{Tseytlin:1990nb}
A.~A. Tseytlin, {\it {Duality symmetric formulation of string world sheet
  dynamics}},  {\em Phys.Lett.} {\bf B242} (1990) 163--174.

\bibitem{Tseytlin:1990va}
A.~A. Tseytlin, {\it {Duality symmetric closed string theory and interacting
  chiral scalars}},  {\em Nucl.Phys.} {\bf B350} (1991) 395--440.

\bibitem{Siegel:1993xq}
W.~Siegel, {\it {Two vierbein formalism for string inspired axionic gravity}},
  {\em Phys.Rev.} {\bf D47} (1993) 5453--5459,
  [\href{http://arxiv.org/abs/hep-th/9302036}{{\tt hep-th/9302036}}].

\bibitem{Siegel:1993th}
W.~Siegel, {\it {Superspace duality in low-energy superstrings}},  {\em
  Phys.Rev.} {\bf D48} (1993) 2826--2837,
  [\href{http://arxiv.org/abs/hep-th/9305073}{{\tt hep-th/9305073}}].

\bibitem{Hohm:2013bwa}
O.~Hohm, D.~L{\"u}st, and B.~Zwiebach, {\it {The Spacetime of Double Field
  Theory: Review, Remarks, and Outlook}},  {\em Fortsch.Phys.} {\bf 61} (2013)
  926--966, [\href{http://arxiv.org/abs/1309.2977}{{\tt arXiv:1309.2977}}].

\bibitem{Kugo:1992md}
T.~Kugo and B.~Zwiebach, {\it {Target space duality as a symmetry of string
  field theory}},  {\em Prog.Theor.Phys.} {\bf 87} (1992) 801--860,
  [\href{http://arxiv.org/abs/hep-th/9201040}{{\tt hep-th/9201040}}].

\bibitem{Hillmann:2009ci}
C.~Hillmann, {\it {Generalized E(7(7)) coset dynamics and D=11 supergravity}},
  {\em JHEP} {\bf 0903} (2009) 135, [\href{http://arxiv.org/abs/0901.1581}{{\tt
  arXiv:0901.1581}}].

\bibitem{Berman:2010is}
D.~S. Berman and M.~J. Perry, {\it {Generalized Geometry and M theory}},  {\em
  JHEP} {\bf 1106} (2011) 074, [\href{http://arxiv.org/abs/1008.1763}{{\tt
  arXiv:1008.1763}}].

\bibitem{Berman:2011cg}
D.~S. Berman, H.~Godazgar, M.~Godazgar, and M.~J. Perry, {\it {The Local
  symmetries of M-theory and their formulation in generalised geometry}},  {\em
  JHEP} {\bf 1201} (2012) 012, [\href{http://arxiv.org/abs/1110.3930}{{\tt
  arXiv:1110.3930}}].

\bibitem{Berman:2011jh}
D.~S. Berman, H.~Godazgar, M.~J. Perry, and P.~West, {\it {Duality Invariant
  Actions and Generalised Geometry}},  {\em JHEP} {\bf 1202} (2012) 108,
  [\href{http://arxiv.org/abs/1111.0459}{{\tt arXiv:1111.0459}}].

\bibitem{Hohm:2013pua}
O.~Hohm and H.~Samtleben, {\it {Exceptional Form of D=11 Supergravity}},  {\em
  Phys.Rev.Lett.} {\bf 111} (2013) 231601,
  [\href{http://arxiv.org/abs/1308.1673}{{\tt arXiv:1308.1673}}].

\bibitem{Park:2013gaj}
J.-H. Park and Y.~Suh, {\it {U-geometry : SL(5)}},  {\em JHEP} {\bf 04} (2013)
  147, [\href{http://arxiv.org/abs/1302.1652}{{\tt arXiv:1302.1652}}].

\bibitem{Aldazabal:2013mya}
G.~Aldazabal, M.~Gra{\~n}a, D.~Marqu{\'e}s, and J.~Rosabal, {\it {Extended
  geometry and gauged maximal supergravity}},  {\em JHEP} {\bf 1306} (2013)
  046, [\href{http://arxiv.org/abs/1302.5419}{{\tt arXiv:1302.5419}}].

\bibitem{Blair:2013gqa}
C.~D.~A. Blair, E.~Malek, and J.-H. Park, {\it {M-theory and Type IIB from a
  Duality Manifest Action}},  {\em JHEP} {\bf 1401} (2014) 172,
  [\href{http://arxiv.org/abs/1311.5109}{{\tt arXiv:1311.5109}}].

\bibitem{Blair:2014zba}
C.~D.~A. Blair and E.~Malek, {\it {Geometry and fluxes of SL(5) exceptional
  field theory}},  {\em JHEP} {\bf 1503} (2015) 144,
  [\href{http://arxiv.org/abs/1412.0635}{{\tt arXiv:1412.0635}}].

\bibitem{Hohm:2014fxa}
O.~Hohm and H.~Samtleben, {\it {Exceptional Field Theory III: E$_{8(8)}$}},
  {\em Phys.Rev.} {\bf D90} (2014) 066002,
  [\href{http://arxiv.org/abs/1406.3348}{{\tt arXiv:1406.3348}}].

\bibitem{Hohm:2013uia}
O.~Hohm and H.~Samtleben, {\it {Exceptional Field Theory II: E$_{7(7)}$}},
  {\em Phys.Rev.} {\bf D89} (2014) 066017,
  [\href{http://arxiv.org/abs/1312.4542}{{\tt arXiv:1312.4542}}].

\bibitem{Hohm:2013vpa}
O.~Hohm and H.~Samtleben, {\it {Exceptional Field Theory I: $E_{6(6)}$
  covariant Form of M-Theory and Type IIB}},  {\em Phys.Rev.} {\bf D89} (2014)
  066016, [\href{http://arxiv.org/abs/1312.0614}{{\tt arXiv:1312.0614}}].

\bibitem{Abzalov:2015ega}
A.~Abzalov, I.~Bakhmatov, and E.~T. Musaev, {\it {Exceptional field theory:
  $SO(5,5)$}},  {\em JHEP} {\bf 06} (2015) 088,
  [\href{http://arxiv.org/abs/1504.01523}{{\tt arXiv:1504.01523}}].

\bibitem{Musaev:2015ces}
E.~T. Musaev, {\it {Exceptional field theory: $SL(5)$}},
  \href{http://arxiv.org/abs/1512.02163}{{\tt arXiv:1512.02163}}.

\bibitem{Hohm:2015xna}
O.~Hohm and Y.-N. Wang, {\it {Tensor Hierarchy and Generalized Cartan Calculus
  in SL(3)$\times$SL(2) Exceptional Field Theory}},
  \href{http://arxiv.org/abs/1501.01600}{{\tt arXiv:1501.01600}}.

\bibitem{Godazgar:2014nqa}
H.~Godazgar, M.~Godazgar, O.~Hohm, H.~Nicolai, and H.~Samtleben, {\it
  {Supersymmetric E$_{7(7)}$ Exceptional Field Theory}},
  \href{http://arxiv.org/abs/1406.3235}{{\tt arXiv:1406.3235}}.

\bibitem{Musaev:2014lna}
E.~Musaev and H.~Samtleben, {\it {Fermions and Supersymmetry in $\rm E_{6(6)}$
  Exceptional Field Theory}},  \href{http://arxiv.org/abs/1412.7286}{{\tt
  arXiv:1412.7286}}.

\bibitem{Cederwall:2013naa}
M.~Cederwall, J.~Edlund, and A.~Karlsson, {\it {Exceptional geometry and tensor
  fields}},  {\em JHEP} {\bf 1307} (2013) 028,
  [\href{http://arxiv.org/abs/1302.6736}{{\tt arXiv:1302.6736}}].

\bibitem{Wang:2015hca}
Y.-N. Wang, {\it {Generalized Cartan Calculus in general dimension}},  {\em
  JHEP} {\bf 07} (2015) 114, [\href{http://arxiv.org/abs/1504.04780}{{\tt
  arXiv:1504.04780}}].

\bibitem{Aldazabal:2013sca}
G.~Aldazabal, D.~Marqu{\'e}s, and C.~N{\'u}{\~n}ez, {\it {Double Field Theory:
  A Pedagogical Review}},  {\em Class.Quant.Grav.} {\bf 30} (2013) 163001,
  [\href{http://arxiv.org/abs/1305.1907}{{\tt arXiv:1305.1907}}].

\bibitem{Berman:2013eva}
D.~S. Berman and D.~C. Thompson, {\it {Duality Symmetric String and M-Theory}},
   {\em Phys. Rept.} {\bf 566} (2014) 1--60,
  [\href{http://arxiv.org/abs/1306.2643}{{\tt arXiv:1306.2643}}].

\bibitem{Choi:2015gia}
K.-S. Choi, {\it {Supergravity in Twelve Dimension}},  {\em JHEP} {\bf 09}
  (2015) 101, [\href{http://arxiv.org/abs/1504.00602}{{\tt arXiv:1504.00602}}].

\bibitem{Linch:2015lwa}
W.~D. Linch and W.~Siegel, {\it {F-theory Superspace}},
  \href{http://arxiv.org/abs/1501.02761}{{\tt arXiv:1501.02761}}.

\bibitem{Linch:2015fya}
W.~D. Linch, III and W.~Siegel, {\it {F-theory from Fundamental Five-branes}},
  \href{http://arxiv.org/abs/1502.00510}{{\tt arXiv:1502.00510}}.

\bibitem{Linch:2015qva}
W.~D. Linch and W.~Siegel, {\it {F-theory with Worldvolume Sectioning}},
  \href{http://arxiv.org/abs/1503.00940}{{\tt arXiv:1503.00940}}.

\bibitem{Linch:2015fca}
W.~D. Linch and W.~Siegel, {\it {Critical Super F-theories}},
  \href{http://arxiv.org/abs/1507.01669}{{\tt arXiv:1507.01669}}.

\bibitem{Siegel:2016dek}
W.~Siegel, {\it {F-theory with zeroth-quantized ghosts}},
  \href{http://arxiv.org/abs/1601.03953}{{\tt arXiv:1601.03953}}.

\bibitem{Schwarz:1995jq}
J.~H. Schwarz, {\it {The power of M theory}},  {\em Phys. Lett.} {\bf B367}
  (1996) 97--103, [\href{http://arxiv.org/abs/hep-th/9510086}{{\tt
  hep-th/9510086}}].

\bibitem{Berman:2014hna}
D.~S. Berman and F.~J. Rudolph, {\it {Strings, Branes and the Self-dual
  Solutions of Exceptional Field Theory}},  {\em JHEP} {\bf 05} (2015) 130,
  [\href{http://arxiv.org/abs/1412.2768}{{\tt arXiv:1412.2768}}].

\bibitem{Aldazabal:2011nj}
G.~Aldazabal, W.~Baron, D.~Marqu{\'e}s, and C.~N{\'u}{\~n}ez, {\it {The
  effective action of Double Field Theory}},  {\em JHEP} {\bf 1111} (2011) 052,
  [\href{http://arxiv.org/abs/1109.0290}{{\tt arXiv:1109.0290}}].

\bibitem{Geissbuhler:2011mx}
D.~Geissb{\"u}hler, {\it {Double Field Theory and N=4 Gauged Supergravity}},
  {\em JHEP} {\bf 1111} (2011) 116, [\href{http://arxiv.org/abs/1109.4280}{{\tt
  arXiv:1109.4280}}].

\bibitem{Grana:2012rr}
M.~Gra{\~n}a and D.~Marqu{\'e}s, {\it {Gauged Double Field Theory}},  {\em
  JHEP} {\bf 1204} (2012) 020, [\href{http://arxiv.org/abs/1201.2924}{{\tt
  arXiv:1201.2924}}].

\bibitem{Berman:2012uy}
D.~S. Berman, E.~T. Musaev, and D.~C. Thompson, {\it {Duality Invariant
  M-theory: Gauged supergravities and Scherk-Schwarz reductions}},  {\em JHEP}
  {\bf 1210} (2012) 174, [\href{http://arxiv.org/abs/1208.0020}{{\tt
  arXiv:1208.0020}}].

\bibitem{deWit:2005hv}
B.~de~Wit and H.~Samtleben, {\it {Gauged maximal supergravities and hierarchies
  of nonAbelian vector-tensor systems}},  {\em Fortsch. Phys.} {\bf 53} (2005)
  442--449, [\href{http://arxiv.org/abs/hep-th/0501243}{{\tt hep-th/0501243}}].

\bibitem{deWit:2008ta}
B.~de~Wit, H.~Nicolai, and H.~Samtleben, {\it {Gauged Supergravities, Tensor
  Hierarchies, and M-Theory}},  {\em JHEP} {\bf 0802} (2008) 044,
  [\href{http://arxiv.org/abs/0801.1294}{{\tt arXiv:0801.1294}}].

\bibitem{Berman:2011kg}
D.~S. Berman, E.~T. Musaev, and M.~J. Perry, {\it {Boundary Terms in
  Generalized Geometry and doubled field theory}},  {\em Phys.Lett.} {\bf B706}
  (2011) 228--231, [\href{http://arxiv.org/abs/1110.3097}{{\tt
  arXiv:1110.3097}}].

\bibitem{Berman:2012vc}
D.~S. Berman, M.~Cederwall, A.~Kleinschmidt, and D.~C. Thompson, {\it {The
  gauge structure of generalised diffeomorphisms}},  {\em JHEP} {\bf 1301}
  (2013) 064, [\href{http://arxiv.org/abs/1208.5884}{{\tt arXiv:1208.5884}}].

\bibitem{Hohm:2012gk}
O.~Hohm and B.~Zwiebach, {\it {Large Gauge Transformations in Double Field
  Theory}},  {\em JHEP} {\bf 1302} (2013) 075,
  [\href{http://arxiv.org/abs/1207.4198}{{\tt arXiv:1207.4198}}].

\bibitem{Park:2013mpa}
J.-H. Park, {\it {Comments on double field theory and diffeomorphisms}},  {\em
  JHEP} {\bf 1306} (2013) 098, [\href{http://arxiv.org/abs/1304.5946}{{\tt
  arXiv:1304.5946}}].

\bibitem{Berman:2014jba}
D.~S. Berman, M.~Cederwall, and M.~J. Perry, {\it {Global aspects of double
  geometry}},  {\em JHEP} {\bf 1409} (2014) 066,
  [\href{http://arxiv.org/abs/1401.1311}{{\tt arXiv:1401.1311}}].

\bibitem{Hull:2014mxa}
C.~M. Hull, {\it {Finite Gauge Transformations and Geometry in Double Field
  Theory}},  {\em JHEP} {\bf 04} (2015) 109,
  [\href{http://arxiv.org/abs/1406.7794}{{\tt arXiv:1406.7794}}].

\bibitem{Naseer:2015tia}
U.~Naseer, {\it {A note on large gauge transformations in double field
  theory}},  {\em JHEP} {\bf 06} (2015) 002,
  [\href{http://arxiv.org/abs/1504.05913}{{\tt arXiv:1504.05913}}].

\bibitem{Rey:2015mba}
S.-J. Rey and Y.~Sakatani, {\it {Finite Transformations in Doubled and
  Exceptional Space}},  \href{http://arxiv.org/abs/1510.06735}{{\tt
  arXiv:1510.06735}}.

\bibitem{Chaemjumrus:2015vap}
N.~Chaemjumrus and C.~M. Hull, {\it {Finite Gauge Transformations and Geometry
  in Extended Field Theory}},  \href{http://arxiv.org/abs/1512.03837}{{\tt
  arXiv:1512.03837}}.

\bibitem{Ortin:2004ms}
T.~Ortin, {\em {Gravity and strings}}.
\newblock Cambridge Univ. Press, 2004.

\bibitem{Baguet:2015xha}
A.~Baguet, O.~Hohm, and H.~Samtleben, {\it {E$_{6(6)}$ Exceptional Field
  Theory: Review and Embedding of Type IIB}},  {\em PoS} {\bf CORFU2014} (2015)
  133, [\href{http://arxiv.org/abs/1506.01065}{{\tt arXiv:1506.01065}}].

\bibitem{Greene:1989ya}
B.~R. Greene, A.~D. Shapere, C.~Vafa, and S.-T. Yau, {\it {Stringy Cosmic
  Strings and Noncompact Calabi-Yau Manifolds}},  {\em Nucl. Phys.} {\bf B337}
  (1990) 1.

\bibitem{Gibbons:1995vg}
G.~W. Gibbons, M.~B. Green, and M.~J. Perry, {\it {Instantons and seven-branes
  in type IIB superstring theory}},  {\em Phys. Lett.} {\bf B370} (1996)
  37--44, [\href{http://arxiv.org/abs/hep-th/9511080}{{\tt hep-th/9511080}}].

\bibitem{Aldazabal:2015yna}
G.~Aldazabal, M.~Graña, S.~Iguri, M.~Mayo, C.~Nuñez, and J.~A. Rosabal, {\it
  {Enhanced gauge symmetry and winding modes in Double Field Theory}},
  \href{http://arxiv.org/abs/1510.07644}{{\tt arXiv:1510.07644}}.

\bibitem{Berkeley:2014nza}
J.~Berkeley, D.~S. Berman, and F.~J. Rudolph, {\it {Strings and Branes are
  Waves}},  {\em JHEP} {\bf 06} (2014) 006,
  [\href{http://arxiv.org/abs/1403.7198}{{\tt arXiv:1403.7198}}].

\bibitem{Blair:2015eba}
C.~D.~A. Blair, {\it {Conserved Currents of Double Field Theory}},
  \href{http://arxiv.org/abs/1507.07541}{{\tt arXiv:1507.07541}}.

\bibitem{Park:2015bza}
J.-H. Park, S.-J. Rey, W.~Rim, and Y.~Sakatani, {\it {O(D,D) Covariant Noether
  Currents and Global Charges in Double Field Theory}},
  \href{http://arxiv.org/abs/1507.07545}{{\tt arXiv:1507.07545}}.

\bibitem{Naseer:2015fba}
U.~Naseer, {\it {Canonical formulation and conserved charges of double field
  theory}},  \href{http://arxiv.org/abs/1508.00844}{{\tt arXiv:1508.00844}}.

\bibitem{Berman:2014jsa}
D.~S. Berman and F.~J. Rudolph, {\it {Branes are Waves and Monopoles}},  {\em
  JHEP} {\bf 05} (2015) 015, [\href{http://arxiv.org/abs/1409.6314}{{\tt
  arXiv:1409.6314}}].

\bibitem{Malek:2015hma}
E.~Malek and H.~Samtleben, {\it {Dualising consistent IIA/IIB truncations}},
  {\em JHEP} {\bf 12} (2015) 029, [\href{http://arxiv.org/abs/1510.03433}{{\tt
  arXiv:1510.03433}}].

\bibitem{Hohm:2011ex}
O.~Hohm and S.~K. Kwak, {\it {Double Field Theory Formulation of Heterotic
  Strings}},  {\em JHEP} {\bf 1106} (2011) 096,
  [\href{http://arxiv.org/abs/1103.2136}{{\tt arXiv:1103.2136}}].

\bibitem{deBoer:2012ma}
J.~de~Boer and M.~Shigemori, {\it {Exotic Branes in String Theory}},  {\em
  Phys.Rept.} {\bf 532} (2013) 65--118,
  [\href{http://arxiv.org/abs/1209.6056}{{\tt arXiv:1209.6056}}].

\bibitem{Braun:2013yla}
A.~P. Braun, F.~Fucito, and J.~F. Morales, {\it {U-folds as K3 fibrations}},
  {\em JHEP} {\bf 10} (2013) 154, [\href{http://arxiv.org/abs/1308.0553}{{\tt
  arXiv:1308.0553}}].

\bibitem{Candelas:2014jma}
P.~Candelas, A.~Constantin, C.~Damian, M.~Larfors, and J.~F. Morales, {\it
  {Type IIB flux vacua from G-theory I}},  {\em JHEP} {\bf 02} (2015) 187,
  [\href{http://arxiv.org/abs/1411.4785}{{\tt arXiv:1411.4785}}].

\bibitem{Candelas:2014kma}
P.~Candelas, A.~Constantin, C.~Damian, M.~Larfors, and J.~F. Morales, {\it
  {Type IIB flux vacua from G-theory II}},  {\em JHEP} {\bf 02} (2015) 188,
  [\href{http://arxiv.org/abs/1411.4786}{{\tt arXiv:1411.4786}}].

\bibitem{Musaev:2013rq}
E.~T. Musaev, {\it {Gauged supergravities in 5 and 6 dimensions from
  generalised Scherk-Schwarz reductions}},  {\em JHEP} {\bf 1305} (2013) 161,
  [\href{http://arxiv.org/abs/1301.0467}{{\tt arXiv:1301.0467}}].

\bibitem{Geissbuhler:2013uka}
D.~Geissb{\"u}hler, D.~Marqu{\'e}s, C.~N{\'u}{\~n}ez, and V.~Penas, {\it
  {Exploring Double Field Theory}},  {\em JHEP} {\bf 1306} (2013) 101,
  [\href{http://arxiv.org/abs/1304.1472}{{\tt arXiv:1304.1472}}].

\bibitem{Berman:2013cli}
D.~S. Berman and K.~Lee, {\it {Supersymmetry for Gauged Double Field Theory and
  Generalised Scherk-Schwarz Reductions}},
  \href{http://arxiv.org/abs/1305.2747}{{\tt arXiv:1305.2747}}.

\bibitem{Condeescu:2013yma}
C.~Condeescu, I.~Florakis, C.~Kounnas, and D.~Lüst, {\it {Gauged
  supergravities and non-geometric Q/R-fluxes from asymmetric orbifold CFT`s}},
   {\em JHEP} {\bf 10} (2013) 057, [\href{http://arxiv.org/abs/1307.0999}{{\tt
  arXiv:1307.0999}}].

\bibitem{Aldazabal:2013via}
G.~Aldazabal, M.~Gra{\~n}a, D.~Marqu{\'e}s, and J.~A. Rosabal, {\it {The gauge
  structure of Exceptional Field Theories and the tensor hierarchy}},  {\em
  JHEP} {\bf 1404} (2014) 049, [\href{http://arxiv.org/abs/1312.4549}{{\tt
  arXiv:1312.4549}}].

\bibitem{Lee:2014mla}
K.~Lee, C.~Strickland-Constable, and D.~Waldram, {\it {Spheres, generalised
  parallelisability and consistent truncations}},
  \href{http://arxiv.org/abs/1401.3360}{{\tt arXiv:1401.3360}}.

\bibitem{Baron:2014yua}
W.~H. Baron, {\it {Gaugings from $E_{7(7)}$ extended geometries}},  {\em Phys.
  Rev.} {\bf D91} (2015), no.~2 024008,
  [\href{http://arxiv.org/abs/1404.7750}{{\tt arXiv:1404.7750}}].

\bibitem{Lee:2015xga}
K.~Lee, C.~Strickland-Constable, and D.~Waldram, {\it {New gaugings and
  non-geometry}},  \href{http://arxiv.org/abs/1506.03457}{{\tt
  arXiv:1506.03457}}.

\end{thebibliography}\endgroup
\end{document}